\documentclass[a4paper,11pt]{article}
\pdfoutput=1
\usepackage{jcappub}

\usepackage{graphicx,subfig}
\usepackage{wrapfig}

\usepackage{amsmath,amssymb,slashed}
\usepackage{tikz}
\usepackage{hyperref}
\usepackage{soul}
\usepackage{tabularx}
\usepackage[normalem]{ulem}

\newcommand{\be}{\begin{equation}}
\newcommand{\ee}{\end{equation}}

\newcommand{\ba}{\begin{eqnarray}}
\newcommand{\ea}{\end{eqnarray}}

\newcommand{\fNL}{f_{\mathrm{NL}}}
\newcommand{\gNL}{g_{\mathrm{NL}}}
\newcommand{\tauNL}{\tau_{\mathrm{NL}}}

\graphicspath{{images/}}

\begin{document}
\title{Constraining general multi-field inflation using the SPHEREx all-sky survey power spectra}

\author[a,b]{Charuhas Shiveshwarkar}
\author[c,d]{Thejs Brinckmann}
\author[a,b]{Marilena Loverde}
\affiliation[a]{C. N. Yang Institute for Theoretical Physics and Department of Physics \& Astronomy, Stony Brook University, Stony Brook, NY 11794, USA}
\affiliation[b]{Department of Physics, University of Washington, Seattle, WA 98195, U.S.A.}

\affiliation[c]{Dipartimento di Fisica e Scienze della Terra, Universit\`a degli Studi di Ferrara, via Giuseppe Saragat 1, I-44122 Ferrara, Italy}
\affiliation[d]{Istituto Nazionale di Fisica Nucleare, Sezione di Ferrara, via Giuseppe Saragat 1, I-44122 Ferrara, Italy}
\emailAdd{Charuhas-Waman.Shiveshwarkar@stonybrook.edu, thejs.brinckmann@unife.it, mloverde@uw.edu}

\abstract{
We investigate how well the SPHEREx all-sky survey can constrain local primordial non-Gaussianity beyond the parameter $\fNL$ using galaxy power spectra. We forecast joint constraints on the parameters $\fNL$, $\gNL$ and $\tauNL$ obtained assuming a simple two-field curvaton model of inflation. The parameters $\fNL$ and $\gNL$ characterise the squeezed limits of the primordial bispectrum and trispectrum respectively, and lead to a characteristic scale-dependence of the galaxy bias that increases out to arbitrarily large scales. Values of the parameter $\tauNL> (\frac{6}{5}\fNL)^2$ cause the galaxy power spectrum to have a stochastic component which also increases out to arbitrarily large scales. Our MCMC forecasts indicate that SPHEREx can provide joint constraints on any two of the three parameters $\fNL,\gNL$ and $\tauNL$. Due to strong degeneracies among these parameters, measurements of the galaxy power spectra alone may not be sufficient to jointly constrain all three. Constraints on $\fNL,\ \gNL$ and $\tauNL$ obtained from galaxy power spectrum observations depend on the modelling of underlying nuisance parameters. We study the robustness of our forecast constraints to modelling choices and note that even with relatively conservative modelling assumptions, SPHEREx galaxy power spectra can provide strong evidence of local non-Gaussianity, even if the particular values of $\fNL$ and $\gNL$ cannot be measured precisely.}

\maketitle

\section{Introduction}
A major goal of cosmology is to determine the source of the initial seeds of structure in our Universe~\cite{Meerburg:2019qqi, Slosar:2019gvt}. In the simplest scenario, the primordial curvature perturbations were sourced by quantum fluctuations in a single inflaton field, whose energy density is also responsible for driving the accelerated expansion of the Universe during inflation~\cite{Guth:1982ec,Starobinsky:1982ee, Bardeen:1983qw}. In this case, the primordial  perturbations are nearly Gaussian, with departures from Gaussianity only arising if the inflaton has strong self-interactions~\cite{Maldacena:2002vr,Chen:2006nt,Cheung:2007st} or unusual scenarios in which the inflaton field is not on the slow-roll attractor solution to the equations of motion \cite{Kinney:2005vj, Martin:2012pe}. If additional light ($m\ll H$, where $H$ is the Hubble rate during inflation) fields are present during inflation, then these fields will acquire a spectrum of primordial fluctuations in addition to the fluctuations from the inflaton. Such fluctuations can be converted into fluctuations in curvature through processes that are generically nonlinear but local in configuration space. Examples of such scenarios are the curvaton model and modulated reheating~\cite{Enqvist:2001zp, Lyth:2001nq, Sasaki:2006kq, Dvali:2003em, Dvali:2003ar}. Curvature perturbations generated in this manner have distinct {\em non-Gaussian} statistical distributions that can be used to infer their presence. 

The non-Gaussian primordial curvature perturbation, $\zeta$, arising from fluctuations in additional light fields can typically be written as,
\begin{equation}
    \label{fNLgNLdef}
    \zeta(\vec{x}) = \mathcal{Z}(\vec{x}) + \frac{3}{5}\fNL(\mathcal{Z}(\vec{x})^2 - \langle \mathcal{Z}^2\rangle)+ \frac{9}{25}\gNL(\mathcal{Z}(\vec{x})^3-3\langle \mathcal{Z}^2\rangle \mathcal{Z}(\vec{x})) + \dots
\end{equation}
where $\mathcal{Z}$ is a Gaussian random field and the coefficients $\fNL$ and $\gNL$ characterise nonlinear terms which are subdominant in the limit $\fNL\langle \mathcal{Z}^2\rangle^{1/2}\,,\gNL \langle \mathcal{Z}^2\rangle \ll1$.\footnote{This definition of $\gNL$ (with the $-3\langle \mathcal{Z}^2\rangle \mathcal{Z}(\vec{x})$ subtracted)  is chosen so that the variance of $\mathcal{Z}$ equals the variance of $\zeta$ at lowest order in $\gNL$.} 
The expansion in Eq.~(\ref{fNLgNLdef}) is said to describe Local Primordial non-Gaussianity (LPnG) because it is generated through a mapping of a Gaussian field that is local in configuration space. The primordial bispectrum of $\zeta$ from Eq.~(\ref{fNLgNLdef}) is proportional to $\fNL$, while the primordial trispectrum contains two terms, one proportional to $\gNL$ and another proportional to $\fNL^2$. A characteristic of LPnG initial conditions is that the higher-order correlation functions of $\zeta$ peak in the so-called squeezed-limit, when one of the wavenumbers of the $\zeta$ fields is taken to zero. This is a feature that is distinct from primordial curvature generated by fluctuations from a single inflaton field~\cite{Creminelli:2004yq}. The detection of LPnG would therefore be of tremendous importance, and imply that the initial conditions of our Universe were sourced by a new light field, over and above the inflaton~\cite{Creminelli:2011rh}. The best constraints on the parameters $\fNL$ and $\gNL$ currently come from the Planck cosmic microwave background (CMB) datasets~\cite{Planck:2019kim}, according to which $\fNL = -0.9\pm 5.1$ and $\gNL = (-5.8\pm6.5)\times 10^4$ at $68\%$ confidence. Typical multi-field inflationary models (including the aforementioned curvaton models) predict $\fNL\approx 1$~\cite{Meerburg:2019qqi}. A robust detection of $\fNL \approx 1$ is therefore an important theoretical target and would conclusively establish the existence of one or more additional light fields during the epoch of inflation. 

A generalisation of the non-Gaussian initial conditions in Eq.~(\ref{fNLgNLdef}) is to allow for an observable contribution to the primordial curvature $\zeta$ that is Gaussian and sourced by the inflaton, in addition to that sourced the curvaton (see, e.g, Refs.~\cite{Shandera:2010ei, Tseliakhovich:2010kf}).\footnote{From this point forward, for simplicity we refer to non-Gaussian initial curvature perturbations as in Eq.~(\ref{fNLgNLdef}) as sourced by the curvaton, though other inflationary models with additional light fields may generate similar initial conditions.}   In this case, the initial curvature perturbations can be written as a sum of terms coming from inflaton, $\zeta_I$ and curvaton, $\zeta_C$,
\begin{eqnarray}
\label{eq:zetaIC}
    \zeta(\vec{x}) = \zeta_{I}(\vec{x})  + \zeta_{C}(\vec{x}) \ .
\end{eqnarray}
Above, $\zeta_I$ is a Gaussian random field and $\zeta_C$ is given by an expansion such as Eq.~(\ref{fNLgNLdef}). As we shall see, these scenarios have suppressed non-Gaussianity, relative to a model where $\zeta_I$ fluctuations are negligible, yet have an $\mathcal{O}(\fNL^2)$ trispectrum that is enhanced relative to the bispectrum in single-source initial conditions from Eq.~(\ref{fNLgNLdef}). This trispectrum peaks in the collapsed limit and its amplitude is characterised by a parameter $\tauNL$ that is bounded from below as $\tauNL \ge (\frac{6}{5}\fNL)^2$~\cite{SuyamaYamaguchi,Smith:2011if}. Initial conditions that exceed this bound can be thought of as having a collapsed trispectrum that is not fully correlated with the square of the primordial bispectrum.  Current bounds on $\tauNL$ from Planck CMB data are $\tauNL \le 2800$ at $95\%$ confidence and therefore far above the minimum threshold set by $\fNL$ constraints. 

Although at present the best constraints on primordial non-Gaussianity come from CMB datasets, major advances are expected to come from large-scale structure~\cite{Meerburg:2019qqi}. For scenarios with LPnG, the squeezed-limit $N+1$-point primordial correlation functions modulate $N$-point functions by the large-scale curvature field $\zeta(\vec{x})$. This in turn manifests as a modulation of the galaxy distribution by $\zeta(\vec{x})$, or a scale-dependent galaxy bias with respect to the density field~\cite{Dalal:2007cu}. The scale-dependent galaxy bias is a key signal (in addition to the galaxy bispectrum) used by large scale structure surveys to constrain $\fNL$~\cite{Slosar:2008hx, Agarwal:2013qta, Giannantonio:2013uqa, Leistedt:2014zqa, Dore:2014cca, Meerburg:2019qqi, Ferraro:2019uce, Cabass:2022ymb, Schlegel:2022vrv, DAmico:2022gki}. Current constraints on $\fNL$ from the BOSS galaxy survey are  $\fNL = 9^{+33}_{-35}$ from the galaxy power spectrum alone and $\fNL = -33\pm 28$ with the addition of the galaxy bispectrum, both at $68\%$ confidence~\cite{Cabass:2022ymb}. Yet, the scale-dependent bias is the combined effect of the squeezed limits of arbitrary higher point primordial correlation functions. In other words, all the non-linear coefficients $\fNL,\ \gNL\,\dots $ in Eq.~(\ref{fNLgNLdef})  each contribute to the scale-dependent galaxy bias (see, e.g, Refs.~\cite{DanielBaumann_2013,KendrickMSmith_2012}). As these non-linear coefficients all contribute to the galaxy power spectrum, there is a strong degeneracy among them when constrained using the galaxy power spectrum alone. On the other hand, as we will discuss, scale-dependent bias constraints can be interpreted as constraints on all higher-order terms appearing in Eq.~(\ref{fNLgNLdef}). The galaxy power spectrum therefore provides broad constraining power on LPnG. The tightest constraints on $\gNL$ from scale-dependent bias come from quasar power spectra, $-4<\gNL\times 10^{-5} <4.7$ at $95\%$ confidence. Joint constraints on $\fNL$, $\gNL$ have been performed in~\cite{Giannantonio:2013uqa, Leistedt:2014zqa}. Relatedly, for initial conditions of the form Eq.~(\ref{eq:zetaIC}), the halo bias also acquires a {\em stochastic} component that is scale-dependent on horizon scales, and has been used to constrain the relative contributions of inflaton and curvaton fluctuations~\cite{Tseliakhovich:2010kf}. Generalising future galaxy power spectrum constraints on primordial non-Gaussianity to include these terms beyond $\fNL$ is important, as information about these higher order statistics may be helpful in distinguishing between different multi-field inflation models, thus helping to further constrain inflationary physics. 

The SPHEREx All-sky galaxy survey aims to achieve constraints on the parameter $\fNL$ at a level $\sigma_{\fNL} \lesssim 0.5$~\cite{Dore:2014cca}. Reaching this threshold would represent an order-of-magnitude improvement over CMB constraints and the first instance of reaching values of $\fNL$ predicted by the simplest two-field inflation scenarios~\cite{Enqvist:2001zp, Lyth:2001nq}. SPHEREx will achieve this target through joint analysis of galaxy auto- and cross-power spectra in combination with galaxy bispectra. While the signatures of $\fNL$, $\gNL$, and $\tauNL$ on the galaxy bispectrum and trispectrum are distinct, owing to the different shapes of their primordial correlation functions, these parameters have nearly common signatures in the galaxy power spectrum. As we will review, both $\fNL$ and $\gNL$ contribute to scale-dependent galaxy bias, while $\tauNL \ge (\frac{6}{5}\fNL)^2$ generates a stochasticity between galaxies and matter that is also probed by galaxy auto- and cross-power spectra. Therefore, the galaxy power spectra measured by SPHEREx can be used to put broad constraints on primordial non-Gaussianity beyond just those set by interpreting data in terms of the $\fNL$ parameter. Our goal in this paper is to forecast constraints achievable by SPHEREx from galaxy power spectra on LPnG where the primordial trispectrum, characterised by $\gNL$ and $\tauNL$, is significant. In this, we will pay special attention to degeneracies among these parameters and possible limitations to interpretation due to uncertainties in modelling their signatures on the galaxy bias. While additional information on these parameters will be available from galaxy bispectra and trispectra, in particular information that could break degeneracies among them, the analysis is considerably more complicated and model-dependent. Galaxy power spectra, therefore, could play an important role in establishing the existence of LPnG, that could subsequently be studied in more detail through higher-order galaxy correlation functions. 

The rest of this paper is organised as follows. In Section~\ref{Theory}, we review the derivation of scale-dependent bias from LPnG, allowing for LPnG that is sourced by arbitrarily high orders in an expansion such as Eq.~(\ref{fNLgNLdef}). In the same section, we review the derivation of scale-dependent stochastic bias sourced by multi-field initial conditions such as Eq.~(\ref{eq:zetaIC}). Here, we also allow the curvaton field to have polynomial terms of arbitrary high order, before specialising to the case where terms after $\gNL \mathcal{Z}^3$ can be neglected. In Section~\ref{sec:method}, we present our likelihood and forecasting pipeline, as well as models we use for the SPHEREx galaxy power spectra. In Section~\ref{sec:results}, we present our forecast constraints on $\fNL$, $\gNL$, and $\tauNL$, along with a study of the robustness of these constraints to bias modelling choices. Our conclusions are presented in Section~\ref{sec:discussion}.

\section{Galaxy bias from local primordial non-Gaussianity}\label{Theory}

In this section, we review the derivation of the effect of LPnG on the galaxy power spectrum measured by galaxy surveys~\cite{Dalal:2007cu}, with particular focus on the impact of higher-order primordial correlation functions, beyond the bispectrum.  Throughout our analysis, we assume a \textit{weak} primordial non-Gaussianity, such that the dimensionless cumulants $\langle\zeta^{N}\rangle_{c}/\langle\zeta^2\rangle^{N/2} \ll 1$, for $N>2$ and where $_c$ refers to the connected part of the correlation function.\footnote{In what follows we use $\langle...\rangle$ and $\langle...\rangle_{c}$ interchangeably to mean connected moments.} We will, however, allow dimensionless cumulants of different orders to be of comparable magnitude, i.e. we will consider examples where $\langle\zeta^M\rangle_{c}/\langle\zeta^2\rangle^{M/2} \sim \langle\zeta^{N}\rangle_c/\langle\zeta^2\rangle^{N/2}$, for $M$, $N>2$ but $M\neq N$. For this to happen with the initial conditions in  Eq.~(\ref{fNLgNLdef}), higher-order coefficients, such as $\gNL$, must be much larger in magnitude than lower-order coefficients, such as $\fNL$. For the bispectrum and trispectum specifically, one would require that $\gNL \sim \fNL A_S^{-1/2}$ for them to have comparable magnitude, where $A_s$ characterises the scalar power spectrum of $\zeta$. Initial conditions such as this can be thought of as having a primordial kurtosis that is small, but as large as the primordial skewness, in comparison to the variance. Non-Gaussianity of this sort is completely consistent with current datasets (and in fact explains the relative magnitude of the current constraints on $\fNL$ and $\gNL$).

In Section~\ref{Theory1}, we review how LPnG leads to a characteristic scale-dependence in the galaxy bias that diverges quadratically at the largest scales. In the following Section~\ref{Theory2}, we review how LPnG gives rise to a stochastic contribution to the galaxy power spectrum in the case of a two-field curvaton model of inflation wherein both the inflaton and the massless curvaton contribute comparably to the primordial curvature perturbation~\cite{Tseliakhovich:2010kf,Smith:2010gx}. In both Sections~\ref{Theory1} and~\ref{Theory2}, we first present a more general result for scale-dependent bias and stochasticity that holds for non-trivial squeezed limits of arbitrary higher point primordial correlation functions and then provide concrete expressions for primordial non-Gaussianity with only a primordial bispectrum and trispectrum. We work exclusively in the comoving synchronous gauge. This has the advantage that it is the unique gauge in which the linear bias relation $(\delta_{g}=b_{g}\delta_{cb})$ remains valid (to linear order) at large scales~\cite{Jeong:2011as}.\footnote{Unlike other gauges where there are additional terms which become important at horizon scales~\cite{Jeong:2011as}.}

\subsection{Scale-dependent galaxy bias from local primordial non-Gaussianity}\label{Theory1}
LPnG is characterised by the primordial curvature perturbation having a non-trivial $N$-point correlation function ($N\geq3$) that peaks in the squeezed limit, i.e. when one of the external momenta becomes soft. It is well understood that to lowest order, the squeezed-limit of a primordial $(N+1)$ correlation function encodes the response of the small-scale $N$-point correlation function in the presence of a long wavelength mode of the field whose momentum becomes soft~\cite{Creminelli:2013poa,Senatore:2012wy}. In particular, for the squeezed-limit of the primordial correlation functions, one can show that for $N\geq 2$, 
\begin{eqnarray}
  \lim_{q\to 0} \frac{\langle\zeta(\vec{q})\zeta(\vec{k_1})\zeta(\vec{k_2})...\zeta(\vec{k_N})\rangle}{P_{\zeta}(q)} = \frac{\partial\langle\zeta(\vec{k_1})...\zeta(\vec{k_N})\rangle}{\partial\zeta_{L}}\ ,\label{softlim1}
\end{eqnarray}
where the right-hand side denotes the variation of the small-scale $N$ point function w.r.t. a constant background curvature perturbation $\zeta_{L}$. The above equation also holds for the linear Cold Dark Matter (CDM) and baryon density contrast $(\delta^{lin}_{cb}$) evaluated at late times, i.e. 
\begin{eqnarray}
  \lim_{q\to 0} \frac{\langle\zeta(\vec{q})\delta^{lin}_{cb}(\vec{k_1})...\delta^{lin}_{cb}(\vec{k_N})\rangle}{P_{\zeta}(q)} = \frac{\partial\langle\delta^{lin}_{cb}(\vec{k_1})...\delta^{lin}_{cb}(\vec{k_N})\rangle}{\partial\zeta_{L}}\ ,\label{softlim2}
\end{eqnarray}
where the linear CDM+baryon density fields are evaluated at typical redshifts probed by galaxy surveys. Under initial conditions with LPnG, the left-hand side of the above equations approaches a non-trivial value implying that the fluctuations of CDM+baryon densities acquire a dependence on an ambient large-scale curvature perturbation $(\zeta_{L})$ that does not vanish even at the largest scales. The equivalence principle forbids such a dependence from being generated from nonlinear gravitational evolution \cite{Creminelli:2011rh}. 

To analyse the effect of LPnG on the halo bias, we adopt the peak-background-split (PBS) approach~\cite{Bardeen:1985tr}. In this approach, the matter density fluctuations can be separated into a large-scale component that does not vary significantly over scales as large as the size of a typical halo and a small scale component which undergoes gravitational collapse to form dark matter haloes. The comoving number density of dark matter tracers, e.g. galaxies or quasars, $n_{g}$ is therefore a function of the connected moments of the linear matter density field $\delta_{cb,R}$ smoothed over some small scale $R$ corresponding to the typical Lagrangian  radius of dark matter haloes.\footnote{While in what follows, we write the tracer density $n_g$ as a function of the smoothed fields $\delta_{cb,R}$, the methodology is more general and only requires a separation of scales between the Fourier modes that produce our tracers and long-wavelength variations in the number density of the tracers $n_g$. } For Gaussian initial conditions, the only non-trivial such moment is the smoothed variance of the linear density field $\sigma^{2}_{R} = \langle\delta^{2}_{cb,R}\rangle$. For non-Gaussian initial conditions, one can have non-trivial higher order moments $S_N(R;z) =\langle\delta^{N}_{cb,R}\rangle $ given by: 
\begin{eqnarray}
    \label{dM_eq}
    S_{N} = \int\prod^{N}_{i=1}\frac{d^{3}\vec{k_i}}{(2\pi)^3}W(k_iR)T_{cb}(k_i,z) \langle\zeta(\vec{k_1})...\zeta(\vec{k_N})\rangle\ ,
\end{eqnarray}
where $W(kR)$ is the window function used to calculate the smoothed matter density field $\delta_{cb,R}$ and $T_{cb}(k,z)$ is the linear transfer function of matter density fluctuations $\delta_{cb}(k,z) = T_{cb}(k,z)\zeta(\vec{k})$. Due to the coupling between small-scale and large-scale modes induced by LPnG following Eq.~\eqref{softlim1}, the moments of smoothed linear matter density fluctuations $S_{N}(R;z)$ acquire a dependence on a background large scale curvature perturbation $\zeta_{L}$, through the dependence of $S_N$ on the squeezed-limit of the primordial $N+1$-point correlation function,\footnote{In comoving coordinates, the squeezed limit can have unphysical terms which arise as a consequence of the slight scale-dependence of primordial fluctuations~\cite{Maldacena:2002vr,Creminelli:2011rh,Senatore:2012wy}. These terms however are a gauge artefact, arising from the choice of how physical separations are separated into the product of a scale factor and comoving separations, and are not observable~\cite{Tanaka:2011aj, Pajer:2013ana} and do not contribute to the scale-dependent bias~\cite{dePutter:2015vga}.}
\begin{eqnarray}
    \label{Snresponse}
    \frac{\partial S_{N}}{\partial \zeta_{L}} &=& \int\prod^{N}_{i=1}\frac{d^{3}\vec{k_i}}{(2\pi)^3}W(k_iR)T_{cb}(k_i,z) \frac{\partial\langle\zeta(\vec{k_1})...\zeta(\vec{k_N})\rangle}{\partial\zeta_{L}}\\
    &=& \int\prod^{N}_{i=1}\frac{d^{3}\vec{k_i}}{(2\pi)^3}W(k_iR)T_{cb}(k_i,z)  \lim_{q\to 0} \frac{\langle\zeta(\vec{q})\zeta(\vec{k_1})\zeta(\vec{k_2})...\zeta(\vec{k_N})\rangle}{P_{\zeta}(q)}\ .
        \label{eq:Snsqueeze}
\end{eqnarray}

The quantity $n_{g}$ will generically depend on the small-scale statistics of the matter field such as its variance and the higher-order cumulants $S_N$. Within a large-scale matter density perturbation $\delta_{L}$, $n_{g}[\delta_{L};\{S_{N}(\zeta_{L})\}]$ acquires an additional dependence on the large scale curvature perturbation $\zeta_{L}$. At linear order in $\zeta_{L}$, we have, 
\be
    \label{eq:deltag}
    n_{g}[\delta_{L};\{S_{i}(\zeta_{L})\}] = n_{g}[0;S^{o}_{i}] + \frac{\partial n_{g}}{\partial\delta_L}[0;S^{o}_{i}]\delta_L + \left(\sum^{\infty}_{N=2}\frac{\partial n_{g}}{\partial S_{N}}\cdot\frac{\partial S_{N}}{\partial\zeta_{L}}\right)\zeta_{L}\ ,
\ee
 where $S^{o}_{i}$ are the moments of the smoothed linear density field in the absence of an ambient large-scale curvature perturbation. For typical models with weak non-Gaussianity, only finitely many terms contribute to the summation in Eq. \eqref{eq:deltag}. Relating $\zeta_{L}$ to $\delta_{L}$ using the transfer function $T_{cb}(k,z)$, one can write a linear bias relation for fluctuations in the abundance of matter density tracers 
\begin{eqnarray}
    \label{biaseq}
    \frac{\delta n_{g}}{n_{g}} = \left(\frac{\partial\log n_{g}}{\partial\delta_{L}} + \left(\sum^{\infty}_{N=2}\frac{\partial\log n_{g}}{\partial S_{N}}\cdot\frac{\partial S_{N}/\partial\zeta_{L}}{T_{cb}(k,z)}\right)\right)\delta_{L}
\end{eqnarray}
We define $b_L = \partial\log n_g/\partial \delta_L$ as the \textit{scale-independent} linear Lagrangian  halo bias which acquires a scale-dependent correction in the presence of LPnG,
\begin{eqnarray}
    \label{deltabL}
    \Delta b_{NG}(k,z) = \left(\sum^{\infty}_{N=2}\frac{\partial\log n_{g}}{\partial S_{N}}\cdot\frac{\partial S_{N}}{\partial\zeta_{L}}\right)\cdot\frac{1}{T_{cb}(k,z)}\ .
\end{eqnarray}
The CDM+Baryon transfer function $T_{cb}(k,z)$ can be obtained by inverting the Poisson equation and is typically written as, 
\begin{eqnarray}
\label{transfer:alternate}
T_{cb}(k,z) &=& \frac{3}{5}\alpha(k,z)\ ,\\
\alpha(k,z) &=& \frac{2k^2 \mathrm{T}(k)D(z)}{3H^{2}_{0}\Omega_{m}}\ ,
\end{eqnarray}
where the function $T(k)$ approaches 1 as $k\rightarrow 0$\footnote{Note that $-(3/5)T(k)$ is the transfer function for the gravitational potential deep in the matter-dominated era.} and $D(z)$ is the linear growth factor normalised to be the scale factor during the matter-dominated era. It is convenient to write the scale-dependent contribution to the bias (Eq.~\eqref{deltabL}) in the following form 
\begin{eqnarray}
    \label{biaseq:alternate}
    \Delta b_{NG}(k,z) = \frac{\mathcal{B}(z)}{\alpha(k,z)}\ ,
\end{eqnarray}
where
\begin{eqnarray}
    \label{generalisedbeta}
    \mathcal{B}(z) = \frac{5}{3}\sum^{\infty}_{N=2}\frac{\partial\log n_{g}}{\partial S_{N}}\cdot\frac{\partial S_{N}}{\partial\zeta_{L}}\ .
\end{eqnarray}
Note that in Eq.~\eqref{generalisedbeta}, $\mathcal{B}(z)$ will depend on properties of the tracer population $n_g$ (for instance, the mass or luminosity of galaxies), but we have suppressed that argument here.
In this way, we see that LPnG induces a scale-dependent galaxy bias that diverges as $1/k^2$ at the largest scales and that its magnitude is parameterised by a weighted sum over the squeezed limits of primordial correlation functions (given by the factors $\partial S_{N}/\partial\zeta_{L}$, see Eqs. \eqref{Snresponse} and \eqref{eq:Snsqueeze}) weighted by the marginal contribution of small-scale statistics of the linear density field to the abundance of dark matter tracers (given by the factors $\partial \log n_{g}/\partial S_{N}$). This fact makes scale-dependent bias a broad probe of any type of local primordial non-Gaussianity. This is in stark contrast to probes such as $N$-point functions  of the CMB, which only constrain $N$-point functions of $\zeta$ at the same order.

We now specialise to the simplest model of LPnG constrained by the CMB datasets -- in this model, the primordial curvature fluctuation can be expressed as a polynomial (at most cubic) in a Gaussian random field as
\be
    \label{zetaNG2}
    \zeta(\vec{x}) = \mathcal{Z}(\vec{x}) + \frac{3}{5}\fNL(\mathcal{Z}(\vec{x})^2 - \langle \mathcal{Z}^2\rangle)+ \frac{9}{25}\gNL(\mathcal{Z}(\vec{x})^3-3\langle \mathcal{Z}^2\rangle \mathcal{Z}(\vec{x}))\ .
\ee
Current constraints on the parameters $\fNL$ and $\gNL$ satisfy the condition for weak non-Gaussianity, i.e. $ \gNL P_{\zeta}, \fNL P_{\zeta}^{1/2} \ll 1$, yet allow for $ \gNL P_{\zeta}\sim \fNL P_{\zeta}^{1/2}$. Under these conditions, the only relevant primordial correlation functions in addition to the power spectrum are the bispectrum and trispectrum, proportional to $\fNL$ and $\gNL$,
\begin{eqnarray}
    \label{Bispectrum}
    B_{\zeta}(\vec{k}_1,\vec{k}_2,\vec{k}_3) &=& \frac{6}{5}\fNL\left(P_{\zeta}(k_1)P_{\zeta}(k_2)+\dots 2 \,\textrm{permutations}\right)\ ,\\
    \label{Trispectrum}
    T_{\zeta}(\vec{k}_1,\vec{k}_2,\vec{k}_3,\vec{k}_4) &=& \frac{54}{25}\gNL\left(P_{\zeta}(k_1)P_{\zeta}(k_2)P_{\zeta}(k_3)+ \dots 3\, \textrm{permutations}  \right)\\
     &&+ \,\frac{36}{25}\fNL^2 \left(P_{\zeta}(k_1)P_{\zeta}(k_2)P_{\zeta}(|\vec{k}_1+\vec{k}_3|)+ 11\, \textrm{permutations}\right)\,.\nonumber
\end{eqnarray}
In our studies of $\gNL$, we are working in the limit that $ \gNL P_{\zeta}\sim \fNL P_{\zeta}^{1/2}$ so that $\fNL \ll \gNL$ and the second term in Eq.~(\ref{Trispectrum}) can be dropped. In the next section, we will generalise to two-field initial conditions, which can enhance the $\fNL^2$ term. The only relevant non-trivial moments of the smoothed linear matter density field ($S_{N}$ defined in Eq. \eqref{dM_eq}) are then proportional to $\fNL$ and $\gNL$ and are as follows,
\begin{eqnarray}
    \label{S2}
    S_{2}(R,z) &=& \sigma^{2}(R,z)\ ,\\
    \label{S3}
    S_{3}(R,z) &=&\frac{3}{5}\fNL\Sigma_{3}(R,z)\ ,\\
    \label{S4}
    S_{4}(R,z) &=&\frac{9}{25}\gNL\Sigma_{4}(R,z)\ ,
\end{eqnarray}
where the factors $\Sigma_3$ and $\Sigma_4$ can be obtained using Eq.~\eqref{dM_eq} and are given by:
\be
    \label{Sig3}
    \Sigma_{3} = 6\int\left(\prod_{i=1}^{2}\frac{d^3 k_i}{(2\pi)^3} W(k_{i}R)T_{cb}(k_i,z)P_{\zeta}(k_i)\right)W(|k_1+k_2|R)T_{cb}(|k_1+k_2|,z)\ ,
    \ee
\be
    \label{Sig4}
    \Sigma_{4} = 24\int\left(\prod_{i=1}^{3}\frac{d^3 k_i}{(2\pi)^3} W(k_{i}R)T_{cb}(k_i,z)P_{\zeta}(k_i)\right) W(|k_1+k_2+k_3|R)T_{cb}(|k_1+k_2+k_3|,z)\ .
\ee
Eqs.~\eqref{Snresponse} and~\eqref{eq:Snsqueeze} show that the response of the moment $S_{N}$ to an ambient long-wavelength curvature perturbation is proportional to the smoothed squeezed limit of the $(N+1)$ linear matter correlation function. Since the only non-trivial linear matter correlation functions (for the model given by Eq.~\eqref{zetaNG2}) are the bispectrum and the trispectrum respectively, it follows that the only non-trivial responses are those of $S_{2}$ and $S_{3}$ (i.e. of the smoothed variance and skewness of the matter density field respectively). From Eqs.~\eqref{Snresponse}, ~\eqref{S2} and~\eqref{S3}, we obtain:
\begin{eqnarray}
    \label{SnResponse1}
    \frac{\partial S_{2}}{\partial\zeta_{L}} &=& \frac{12}{5}\fNL S_2\ ,\\
    \label{SnResponse11}
    \frac{\partial S_{3}}{\partial\zeta_{L}} &=& \frac{27}{25}\gNL \Sigma_3\ ,
\end{eqnarray}
where $\Sigma_{3}$ is given by Eq.~\eqref{Sig3}. Substituting Eqs.~\eqref{SnResponse1} and~\eqref{SnResponse11} in Eq.~\eqref{generalisedbeta}, we obtain the following equation for the scale-dependent bias in this simple LPnG model:
\be
    \label{biaseqfNLgNL}
   \left. \mathcal{B}(z) \right|_{\fNL,\gNL}= 4\fNL S_{2}\cdot\frac{\partial \log n_{g}}{\partial S_{2}} + \frac{9}{5}\gNL\Sigma_{3}\cdot\frac{\partial\log n_{g}}{\partial S_{3}}\ .
\ee
Using Eq.~\eqref{S3}, we can express $\Sigma_{3}\left(\partial\log n_{g}/\partial S_{3}\right) = (5/3)\times \partial\log n_{g}/\partial\fNL$ and thus rewrite the non-Gaussian contribution to the galaxy bias in its most often quoted form:

\begin{eqnarray}
    \label{dbNGfNLgNL1}
    \Delta b_{\text{NG}} = \frac{\fNL\beta_{f}(z) + \gNL\beta_{g}(z)}{\alpha(k,z)}\ ,
\end{eqnarray}
where the quantities $\beta_{f}(z) = 2\partial\log n_{g}/\partial\log\sigma$ and $\beta_{g}(z) = 3\partial\log n_{g}/\partial\fNL$ denote the responses of the abundance of matter density tracers to the variance and skewness of the smoothed matter density field, respectively. They are dependent on the complicated physics of galaxy (or other tracer) formation and may not be expressible analytically. 

 In a galaxy power spectrum forecast, we need to make modelling choices for the parameters $\beta_{f}$ and $\beta_{g}$. A strong modelling choice is to assume that the mean tracer density $n_{g}$ has the form of a universal halo mass function -- wherein $n_{g}$ is only a function of the ratio $\delta_{c}/\sigma$, where $\delta_{c}$ is the threshold density for spherical collapse in an Einstein-deSitter universe. In this case, one can show that $\beta_{f} = 2 b_L \delta_c$. Modelling $\beta_{g}$ requires further assumptions about how the tracer density $n_{g}$ depends on $\fNL$. For our SPHEREx forecasts (under strong modelling assumptions), we use the following fitting formula derived in Ref.~\cite{KendrickMSmith_2012} (see their Eq.~(51)):
\begin{eqnarray}
    \label{betagnl}
    \beta_g = \kappa_{3}\left(-0.4(\nu-1)+1.5(\nu-1)^{2}+0.6(\nu-1)^{2}\right)\ ,
\end{eqnarray}
where $\kappa_{3} = 0.000329(1+0.09z)b_{L}^{-0.09}$ and $\nu=(1+0.5\beta_{f})^{1/2}$.

The universality assumption has been widely used to model the scale-dependent galaxy bias induced by LPnG. However, hydrodynamical separate universe simulations (see Refs.~\cite{Barreira_2020,Barreira_2022}) have shown that the universality assumption may not hold true for realistic galaxy populations. In particular, Ref.~\cite{Barreira_2020} has shown that for galaxies selected by stellar mass, $\beta_{f}$ can be modelled more accurately as $\beta_{f} = 2\delta_{c}(b_{E}-p)$, where $0.4\leq p\leq 0.7$, and $b_E$ is the (scale-independent) Eulerian bias. Inspired by this result, we also obtain MCMC forecasts with a non-universal modelling choice whereby $\beta_{f}=2\delta_c(b_{E}-0.5)$ and $\beta_{g}$ is given by Eq.~\eqref{betagnl}, with $\nu = (1+\delta_{c}(b_{E}-0.5))^{1/2}$. In Sec.~\ref{ssec:modelchoices}, we study the robustness of our forecast constraints to these choices by allowing $p$ to be a free parameter.

If $\fNL$ and $\gNL$ are of comparable magnitude, then our expressions for $\beta_{g}$ and $\beta_{f}$ show that the contribution of $\gNL$ to the scale-dependent bias signal is suppressed w.r.t. contribution of $\fNL$ by four orders of magnitude. In this scenario, the primordial correlation functions scale hierarchically, e.g. $S_{N+1} \sim S_N \sigma$, and scale-dependent bias is primarily probing the amplitude of the largest non-trivial cumulant, $\fNL$. In our forecasts with non-zero $\gNL$ we focus on the opposite limit, in which $\gNL \sim 10^4\fNL$, leading to a primordial kurtosis of comparable magnitude to the primordial skewness, $S_3\sim S_4$, so that both terms contribute comparably to the scale-dependent bias. This situation is consistent with current constraints on both parameters and is also the only scenario in which $\gNL$ can be detected in the foreseeable future.

As $b_E$ is typically taken as a free parameter for each tracer population, Eq.~\eqref{dbNGfNLgNL1} shows that the parameters $\fNL$ and $\gNL$ are degenerate with respect to their impact on the scale-dependent bias. In fact, for a single redshift and single tracer, $\fNL$ and $(\beta_{g}/\beta_{f})\gNL$ cannot be distinguished from each other by galaxy power spectrum measurements alone. However, observations of multiple tracer populations, which typically have different values of $\beta_{g}(z)$ and $\beta_{f}(z)$, may break this near-perfect degeneracy. In this paper, we perform MCMC forecasts to quantify the possible degradation in the joint constraints on both $\fNL$ and $\gNL$ that can be obtained from the data collected by the SPHEREx all-sky survey~\cite{Dore:2014cca}.

\subsection{Stochastic galaxy bias from local primordial non-Gaussianity}\label{Theory2}

As stated before, any model of cosmic inflation that produces LPnG in the primordial curvature perturbations necessarily needs to include at least one light field in addition to the inflaton. A simple model that can produce non-Gaussian primordial fluctuations compatible with current measurements of $\fNL$ and $\gNL$ incorporates a single massless field called the \textit{curvaton}~\cite{Enqvist:2001zp,Lyth:2001nq, Sasaki:2006kq} in addition to the inflaton, such that the inflationary expansion of the universe is driven mainly by the dynamics of the inflaton, whereas the primordial fluctuations are completely dominated by fluctuations in the curvaton. If both the inflaton and the curvaton contribute comparably to generating the primordial curvature perturbations, the primordial statistics are changed. In the simplest scenario, the contribution of the inflaton remains Gaussian, while the curvaton contribution remains non-Gaussian and admits an expansion like Eq. \eqref{zetaNG2}, which is local in real space in terms of a Gaussian random variable, i.e.:
\begin{eqnarray}
    \label{zetaNGtauNL}
    \zeta(\vec{x}) &=& \zeta_{I}(\vec{x}) + \zeta_{C}(\vec{x})\ ,\\
       \label{eq:zetaCexp}
    \zeta_{C}(\vec{x}) &=& \mathcal{Z}_{C}(\vec{x}) + \frac{3}{5}\overline{\fNL}(\mathcal{Z}_{C}(\vec{x})^2 - \langle \mathcal{Z}_{C}^2\rangle)+ \frac{9}{25}\overline{\gNL}(\mathcal{Z}_{C}(\vec{x})^3-3\langle \mathcal{Z}_{C}^2\rangle \mathcal{Z}(\vec{x})) + \dots\ , 
\end{eqnarray}
where $\zeta_{I}$ is the Gaussian contribution to the primordial curvature perturbation that is sourced from the inflaton fluctuations, whereas the non-Gaussian $\zeta_{C}$ is sourced from the fluctuations of the curvaton. The quantities $ \overline{\fNL}$ and $\overline{\gNL}$ characterise the amplitudes of the bispectrum and trispectrum of the $\zeta_C$ field. 

The correlation between the fields $\zeta_{I}$ and $\zeta_{C}$ is dependent on the particular features of the inflationary model. For this work, we consider a simple model in which $\zeta_{I}$ and $\zeta_{C}$ are uncorrelated and have proportional power spectra. We define the contribution of the inflaton to the primordial curvature perturbation w.r.t. the curvaton contribution by 
\be
\label{eq:xidef}
\xi^2 = P_{\zeta_I}/P_{\zeta_C}\,,
\ee
a scale-independent constant (for generalisations of this, see, e.g, Ref.~\cite{Shandera:2010ei}). Assuming this simple model, the bispectrum and trispectrum can be computed as,
\ba
    \label{eq:bCI}
    B_{\zeta}(\vec{k}_1,\vec{k}_2,\vec{k}_3) &=& \frac{6}{5}\frac{\overline{\fNL}}{(1+\xi^2)^2}\left(P_{\zeta}(k_1)P_{\zeta}(k_2)+\dots 2 \,\textrm{permutations}\right)\ ,\\
    \label{eq:triCI}
    T_{\zeta}(\vec{k}_1,\vec{k}_2,\vec{k}_3,\vec{k}_4) &=& \frac{54}{25}\frac{\overline{\gNL}}{(1+\xi^2)^3}\left(P_{\zeta}(k_1)P_{\zeta}(k_2)P_{\zeta}(k_3)+ \dots 3\, \textrm{permutations}  \right)\\
    &&+ \,\frac{36}{25}\frac{\overline{\fNL}^2}{(1+\xi^2)^3}\left(P_{\zeta}(k_1)P_{\zeta}(k_2)P_{\zeta}(|\vec{k}_1+\vec{k}_3|)+ 11\, \textrm{permutations}\right)\ .\nonumber
\ea
From above, we identify the parameters $\fNL$ and $\gNL$ that characterize the bispectrum and trispectrum of $\zeta$ as, 
\begin{eqnarray}
    \label{PNGredef}
   \fNL &=&\frac{  \overline{\fNL}}{(1+\xi^2)^2}\ ,\\
    \gNL &=& \frac{\overline{\gNL} }{ (1+\xi^2)^3}\ .
\end{eqnarray}
We see then that for a given level of non-Gaussianity in the $\zeta_C$ field, allowing $\zeta$ to have a contribution from Gaussian inflaton fluctuations reduces the overall amplitude of non-Gaussianity by factors of $P_{\zeta_C}/P_\zeta = 1/(1+\xi^2)$. On the other hand, the collapsed ($\mathcal{O}(\fNL^2)$) part of the trispectrum is enhanced relative to the bispectrum by a factor of $(1+\xi^2)$. Defining
\ba
\label{eq:tauNLdef}
\tauNL &=&  \frac{36}{25}\frac{\overline{\fNL}^2}{(1+\xi^2)^3}\\
 &=& \left(\frac{6}{5}\fNL\right)^{2}(1+\xi^2)\ ,
\ea
we see that $\tauNL$ is enhanced by contributions from the inflaton~\cite{SuyamaYamaguchi}. The most recent Planck datasets~\cite{Planck2013} lead to the constraint $\tauNL< 2800$ at $95\%$ confidence -- this constraint is consistent with the assumption of weak primordial non-Gaussianity ($\langle\zeta^N\rangle/\langle\zeta^2\rangle^{N/2} \ll 1$). Under these conditions, the only non-trivial higher point functions of the primordial curvature perturbations (Eq. \eqref{zetaNGtauNL}) are the bispectrum and the trispectrum of $\zeta_C$, which have the functional forms given by Eqs.~\eqref{Bispectrum} and~\eqref{Trispectrum}, respectively.

To analyse the impact of a primordial curvature perturbation of the kind given by Eq. \eqref{zetaNGtauNL} on the galaxy number density fluctuations, we note that small-scale correlation functions of the curvature perturbation $\zeta$ couples only to an ambient long-wavelength mode of the non-Gaussian part of primordial curvature perturbation, i.e. $\zeta_{C}$. In particular, Eq. \eqref{softlim1} holds in a slightly modified form where now only the non-Gaussian contribution to the curvature, i.e. $\zeta_C$, modulates the small-scale correlation functions,
\be
    \label{softlim1a}
    \lim_{q\to 0} \frac{\langle\zeta_{C}(\vec{q})\zeta(\vec{k_1})...\zeta(\vec{k_N})\rangle}{P_{\zeta_{C}}(q)} = \frac{\partial\langle\zeta(\vec{k_1})...\zeta(\vec{k_N})\rangle}{\partial\zeta_{C,L}}\ , 
\ee
whereas for $N\geq 2$ the curvature perturbations sourced by the inflaton do not modulate small-scale correlation functions,
\ba
    \label{softlim1b}
     \frac{\partial\langle\zeta(\vec{k_1})..\zeta(\vec{k_N})\rangle}{\partial\zeta_{I,L}} &=& \lim_{q\to 0}\frac{\langle\zeta_{I}(\vec{q})\zeta(\vec{k_1})...\zeta(\vec{k_N})\rangle}{P_{\zeta_{I}}(q)} \\
     &=& 0\ .\nonumber
\ea
The moments of the smoothed small-scale linear matter density field $S_{N}(R,z)$ (defined in Eq.~\eqref{dM_eq}) in the presence of a long-wavelength curvature perturbation $\zeta_{L}=\zeta_{I,L}+\zeta_{C,L}$ depend on $\zeta_{C,L}$ similarly as Eq. \eqref{Snresponse},
\be
    \label{Snresponse1a}
      \frac{\partial S_{N}}{\partial \zeta_{C,L}} = \int\prod^{N}_{i=1}\frac{d^{3}\vec{k_i}}{(2\pi)^3}W(k_iR)T_{cb}(k_i,z)  \lim_{q\to 0} \frac{\langle\zeta_{C}(\vec{q})\zeta(\vec{k_1})...\zeta(\vec{k_N})\rangle}{P_{\zeta_{C}}(q)}\ .
\ee

From the arguments of the previous section, it follows therefore that the galaxy number density fluctuations obey a relation similar to Eq. \eqref{biaseq}, namely
\begin{eqnarray}
    \label{biaseq2}
    \frac{\delta n_{g}}{n_{g}} = \frac{\partial\log n_{g}}{\partial\delta_{L}}\delta_{L} + \left(\sum^{\infty}_{N=2}\frac{\partial\log n_{g}}{\partial S_{N}}\cdot\frac{\partial S_{N}}{\partial\zeta_{C,L}}\right)\zeta_{C,L}\ ,
\end{eqnarray}
where $\delta_{L}=T_{cb}(k,z)\zeta_L(\vec{k})=T_{cb}(k,z)(\zeta_{I,L} + \zeta_{C,L})$ is the large scale (linear) density perturbation. Eq. \eqref{biaseq2} shows that the galaxy number density contrast $\delta_{g}=\delta n_{g}/n_{g}$ is not fully correlated with the ambient large scale matter density field $\delta_L$ -- it is also partially correlated with $\zeta_{C,L}$, which has fluctuations that cannot be completely specified from the realisation of $\zeta$ or $\delta_L$ alone. This introduces stochasticity in the distribution of galaxies, relative to the matter field, and is known as \textit{stochastic} bias~\cite{Tseliakhovich:2010kf}. From Eq. \eqref{biaseq2}, the galaxy bias $b_{g}$ and the galaxy power spectrum $P_{gg}$ are given by 
\begin{eqnarray}
    b_{g}(k,z) &=& \frac{\langle \delta_{g}\delta_{L}\rangle}{\langle\delta_{L}\delta_{L}\rangle}\\
    &=& \frac{\partial\log n_{g}}{\partial\delta_{L}} + \frac{1}{1+\xi^2} \left(\sum^{\infty}_{N=2}\frac{\partial\log n_{g}}{\partial S_{N}}\cdot\frac{\partial S_{N}}{\partial\zeta_{C,L}}\right)\frac{1}{T_{cb}(k,z)}\ ,
\end{eqnarray}
where, as before, $b_{L}=\partial\log n_{g}/\partial\delta_{L}$ is the scale-independent, Lagrangian bias, while the non-Gaussian, scale-dependent correction to the galaxy bias is
\begin{eqnarray}
\label{eq:deltabNLC}
    \Delta b_{NG} = \frac{\mathcal{B}_{C}(z)}{\alpha(k,z)}\ ,
\end{eqnarray}
where $\alpha(k,z)$ is defined in the previous section (Eq. \eqref{transfer:alternate}) and 
\begin{eqnarray}
    \label{generalisedbetaC}
    \mathcal{B}_{C}(z) = \frac{5/3}{(1+\xi^2)}\sum^{\infty}_{N=2}\frac{\partial\log n_{g}}{\partial S_{N}}\cdot\frac{\partial S_{N}}{\partial\zeta_{C,L}}\ . 
\end{eqnarray}
Equation \eqref{biaseq2} can now be written in a concise form to give the modified linear bias relation for (Lagrangian) galaxy number density contrasts,
\begin{eqnarray}
    \label{biaseqC}
    \delta_{g} = b_{L}\delta_{L} + (1+\xi^2)\Delta b_{NG}(k,z)\delta_{C,L}\ ,
\end{eqnarray}
where $\delta_{C,L} = T_{cb}(k,z)\zeta_{C,L}$ is the part of the density field sourced by $\zeta_C$. The only non-trivial responses of the smoothed moments of the density field (for the model given by Eqs.~\eqref{zetaNGtauNL} and~\eqref{eq:zetaCexp}) can be obtained using Eq.~\eqref{Snresponse1a} and are given by:
\begin{eqnarray}
    \label{S2ResponseXi2}
    \frac{\partial S_{2}}{\partial\zeta_{L}} &=& \frac{12}{5}\fNL (1+\xi^2) S_{2}\ ,\\
    \label{S3ResponseXi2}
    \frac{\partial S_{3}}{\partial\zeta_{L}} &=& \frac{27}{25}\gNL (1+\xi^2) \Sigma_{3}\ ,
\end{eqnarray}
where $\Sigma_{3}$ is given by Eq.~\eqref{Sig3}. Substituting Eqs.~\eqref{S2ResponseXi2} and~\eqref{S3ResponseXi2} in Eq.~\eqref{generalisedbetaC}, we see that the non-Gaussian correction to the galaxy bias $\Delta b_{NG}$ in the two-field initial conditions given by Eq.~\eqref{zetaNGtauNL} is the same as Eq. \eqref{dbNGfNLgNL1}. The galaxy power spectrum $P_{gg}=\langle\delta_g\delta_g\rangle$ on the other hand can be shown to be,\footnote{Note that $\langle\delta_{C,L}\delta_{C,L}\rangle \propto \langle\delta_{L}\delta_{L}\rangle/(1+\xi^2)$.} 
\begin{eqnarray}
    \label{PggtauNL}
    P_{gg}(k,z)=\left(\left(b_{E}+\Delta b_{NG}\right)^2 + \xi^2\Delta b_{NG}^2\right) P_{cc}(k,z)\ ,
\end{eqnarray}
where $b_{E}=1+b_{L}$ is the Eulerian galaxy bias, $P_{cc}(k,z)$ is the linear CDM+Baryon power spectrum, and the reader is reminded that $\xi^2$ represents the ratio of inflaton-to-curvaton power, Eq.~(\ref{eq:xidef}), so that $\tauNL = (\frac{6}{5}\fNL)^2 (1+\xi^2)$. Equation \eqref{PggtauNL} is the main result of this section, which shows that the galaxy power spectrum acquires a stochastic contribution proportional to the relative contribution of the inflaton to the primordial curvature perturbation w.r.t. the curvaton. We note here an interesting feature of this scenario is that  the different $k$-dependences of the $\Delta b_{NG}$ and $\Delta b_{NG}^2$ terms in Eq.~(\ref{PggtauNL}) allow the combination $(1+\xi^2)$ to be extracted from $P_{gg}$ even if the coefficient $\mathcal{B_C}$ in Eqs.~(\ref{eq:deltabNLC}) \&~(\ref{generalisedbetaC}) cannot be determined, for instance due to the challenges associated with modelling $n_g(S_N)$. 

For the stochastic bias signal to be large enough for a viable measurement of $\tauNL$, we need the stochastic part of the galaxy power spectrum to be comparable to its non-stochastic part. In other words, a robust measurement of $\tauNL$ can be made using the stochastic halo bias if (following Eq.~\eqref{PggtauNL}), $\xi^2 \Delta b_{NG}/(b_{E}+\Delta b_{NG})^2 \sim \mathcal{O}(1)$ in some wavelength and redshift bin which has a significant constraining power. Our choices of the fiducial values for $\tauNL$ will be motivated by imposing this requirement at $z=0$ and at the largest scales accessible to the SPHEREx survey (i.e. $k_{\text{min}} = 0.001\ \text{h/Mpc}$). 

In general, the parameter $\tauNL$, which parameterises the collapsed limit of the primordial 4-point function, can be shown to satisfy the Suyama-Yamaguchi inequality~\cite{SuyamaYamaguchi, Smith:2011if}: $\tauNL\geq(\frac{6}{5}\fNL)^2$. It can also be shown (see, for example, Ref.~\cite{DanielBaumann_2013}) that a significant stochasticity in the galaxy bias arises whenever this inequality is strictly obeyed $\tauNL > (\frac{6}{5}\fNL)^2$.\footnote{In general, it can be shown that the collapsed limits of \textit{all} higher point primordial correlation function leads to stochasticity in the distribution of galaxies~\cite{DanielBaumann_2013}; somewhat analogous to how squeezed limits of all higher point primordial correlations contribute to the scale-dependent bias. However, typical inflationary models constrained by data suggest that the dominant contribution to halo-stochasticity comes from the collapsed limit of the primordial 4-point function whose magnitude is given by $\tauNL$~\cite{DanielBaumann_2013}. Reflecting this fact, we choose to forecast measurements of $\tauNL$ and not other collapsed-limit non-Gaussianity parameters in this paper.} In this section, we have shown how this is true for a simple model\footnote{Note that even a non-trivial $\gNL$ can also lead to stochasticity in the galaxy bias, but this stochasticity is proportional to $P_{\zeta}^2$ and is smaller than the stochasticity due to $\tauNL$ by a factor of $P_{\zeta}\sim 10^{-9}$ and can be safely ignored in the leading-order calculation relevant for this paper. See, for example, Ref.~\cite{DanielBaumann_2013} for a more detailed analysis on how primordial non-Gaussianity can lead to stochastic galaxy bias.} (given by Eq. \eqref{zetaNGtauNL}), which we use to obtain forecasts for $\tauNL$ for the SPHEREx all-sky survey.

\section{Analysis Method}
\label{sec:method}

In this paper, we use the Bayesian likelihood pipeline developed in~\cite{Shiveshwarkar:2023xjv} to forecast the sensitivity of the SPHEREx all-sky survey in jointly constraining the LPnG parameters $\fNL,\ \gNL\ \text{and}\ \tauNL$ around fiducial values consistent with state-of-the-art constraints from CMB datasets. This is a standard Bayesian Markov Chain Monte Carlo (MCMC) method used to obtain parameter forecasts for cosmological surveys -- in particular, we use a multitracer galaxy power spectrum likelihood (developed in Ref.~\cite{Shiveshwarkar:2023xjv}) for a SPHEREx-like galaxy survey constructed within the MontePython v3.4 MCMC sampling package~\cite{Audren:2012wb,MontePython3}.  In this section, we give a broad overview of our MCMC analysis method used for the forecasts obtained in this paper and refer the reader to Ref.~\cite{Shiveshwarkar:2023xjv} for more details.\\
\\
Our MCMC analysis involves the following steps:
\begin{itemize}
    \item Modelling the observables (in this case the galaxy auto- and cross-power spectra at different scales and redshifts) for a given set of cosmological parameters.
    \item Obtaining the posterior likelihood for a given set of cosmological parameters (using the modelling framework defined above) given a `mock data' provided by observables modelled in the fiducial cosmology.
    \item Obtaining confidence intervals for the cosmological parameters by sampling the posterior likelihood function using Markov Chain Monte Carlo (MCMC) methods. Alternatively, one could obtain (potentially) less accurate confidence intervals by computing a Fisher matrix. In this paper, unless otherwise specified, we sample the posterior likelihood function using the Metropolis Hastings MCMC sampling algorithm within the MontePython v3.4 MCMC sampling package.
\end{itemize}
Obtained in this way, the mean cosmological parameters should be the same (within numerical uncertainties) as the fiducial parameters, whereas the confidence intervals provide the forecast sensitivity to the cosmological parameters. 

The SPHEREx all-sky survey~\cite{Dore:2014cca} is designed to minimise systematics in observations of galaxy clustering at large scales (going to $k_{\text{min}}=0.001\ \text{h/Mpc}$). It will map a large cosmic volume and measure spectroscopic redshifts of galaxies belonging to the all-sky catalogues of the WISE~\cite{WISE}, Pan-STARRS~\cite{PanSTARRS} and DES~\cite{DES} galaxy surveys. It is also a deep survey -- observing galaxies in eleven redshift bins $\left[z_{i}^{\text{min}},z_{i}^{\text{max}}\right]_{i=1}^{i=11}$ with mean redshifts $\overline{z}_{i}$ going from $\overline{z}_{\text{min}}=0.1$ to $\overline{z}_{\text{max}}=4.3$.  Due to its low spectral resolution (as compared to a true spectroscopic galaxy survey), SPHEREx is expected to measure spectroscopic galaxy redshifts to varying degrees of precision and will divide its observed galaxies into five galaxy populations according to their redshift uncertainties. In this way, the data collected by SPHEREx will lend itself to the use of multitracer techniques which can be used to obtain significantly tighter constraints on primordial non-Gaussianity. The observables to be obtained from SPHEREx are the galaxy power spectra and galaxy bispectra in redshift space -- in this paper, we study the constraining power of the SPHEREx all-sky survey in constraining LPnG parameters using the galaxy power spectra alone. As discussed, galaxy power spectra alone could provide a first detection of {\em any} LPnG that could later be studied in more detail through higher-order correlation functions. Note that since SPHEREx will divide its observed galaxies into five galaxy populations, our observables include the cross-power spectra as well the auto-power spectra of different galaxy samples. 

As the first step in our likelihood analysis, we need to model the galaxy power spectra for a given set of cosmological parameters. To do so, we use the linear bias relation ($\delta_{g} = b_{g}\delta_{cb}$) that is valid (at the linear level) at late times and in the comoving synchronous gauge (see Eq. \eqref{PggtauNL}). According to the linear bias model, the theoretical galaxy power spectrum in the presence of LPnG between the $i^{th}$ and $j^{th}$ SPHEREx galaxy samples is given by  
\begin{eqnarray}
\label{Pgg_theory}
    P^{th}_{ij}(k,z) &=& b_{g,i}b_{g,j} P_{cc}(k,z) +\xi^{2}\Delta b_{NG,i}\Delta b_{NG,j} P_{cc}(k,z)\ ,
\end{eqnarray}
where $b_{g,i}$ and $\Delta b_{NG,i}$ are respectively the total galaxy bias (including non-Gaussian corrections) and the non-Gaussian contribution (due to both $\fNL$ and $\gNL$) to the galaxy bias of the $i^{th}$ SPHEREx galaxy population. As mentioned in the previous section, $P_{cc}$ is the linear CDM+Baryon power spectrum. The parameter $\xi^2$ is defined such that $\tauNL = (6/5\fNL)^2(1+\xi^2)$ and encodes the relative contribution of the inflaton w.r.t. the curvaton in the primordial curvature fluctuations. As shown in Ref.~\cite{Shiveshwarkar:2023xjv} the sensitivity of SPHEREx to $\fNL$ is entirely driven by low-$k$ Fourier modes where use of the linear power spectrum and linear bias factors are justified. Since the $k$-dependence of the bias is identical for $\fNL$, $\gNL$ and $\tauNL$, this is also justified here. We use the Boltzmann code CLASS v3.2.0~\cite{Blas:2011rf,Lesgourgues:2011re,Lesgourgues:2011rh} to compute the linear CDM+Baryon power spectrum $P_{cc}(k,z)$, and model the non-Gaussian correction $\Delta b_{NG}$ to the galaxy bias according to the analysis of Section~\ref{Theory} -- in particular following Eq.~\eqref{dbNGfNLgNL1} for a cosmology with non-zero $\fNL$ and $\gNL$ and Eq.~(\ref{PggtauNL}) for $\tauNL \neq (\frac{6}{5}\fNL)^2$.

For the galaxy power spectrum forecasts obtained in this paper, we model the nuisance parameter $\beta_{f}$ as 
\be
\label{eq:betafstrong}
\beta_{f} = 2\delta_{c}(b_{E}-p)\,.
\ee
Our first sets of forecasts will fix either $p=1$ and $p=0.5$ with $b_E$ a free parameter for each galaxy sample and each redshift. We subsequently explore consequences of treating $p$ as a free parameter for each SPHEREx galaxy sample. The choice $p=1$ holds in the particular case when the mean number density of observed galaxies has the same form as a universal halo mass function. If this universality assumption does not hold true, the value of $\beta_{f}$ (which in general is tracer-dependent) is difficult to estimate from first principles as it encodes all the intractable details of the non-linear physics of galaxy formation. However, the results of hydrodynamical simulations within the separate universe framework conducted in Refs.~\cite{Barreira_2020,Barreira_2022} suggest that $\beta_{f}$ is related to the galaxy bias as $\beta_{f}=(b_{E}-p)\delta_{c}$ with a tracer-dependent number $p$. Our choice of $p=0.5$ is representative of modelling for $\beta_{f}$ following the observations in Refs.~\cite{Barreira_2020,Barreira_2022} that for galaxies selected by stellar mass, $0.4\leq p \leq 0.7$ is a better fit to the results of hydrodynamical separate universe simulations. SPHEREx galaxy samples are not selected by stellar mass, so we forecast two choices of $p$ to illustrate the $p$-dependence of constraints. 

For modelling $\beta_{g}$, we use the fitting formula Eq.~\eqref{betagnl} derived in Ref.~\cite{KendrickMSmith_2012} which relates $\beta_{g}$ to $\beta_{f}$ through $\nu=(1+0.5\beta_{f})^{1/2}$, with $\beta_f$ as in Eq.~(\ref{eq:betafstrong}). Note that our first-round forecasts are obtained using these strong modelling choices that assume a perfect knowledge of the functional dependence of the nuisance parameters $\beta_{f}$ and $\beta_{g}$ on $b_E$. We also study forecasts under a more conservative modelling choice where $p$ a free, but redshift-independent, parameter. If $p$ were completely free (i.e. allowed to vary independently at different redshifts), then in the limit of large uncertainty on $p$, $\beta_f$ and $\beta_g$ would become free-parameters independent of $b_E$. In the case where there is large uncertainty on the form of $\beta_f$ and $\beta_g$, $\fNL$ and $\gNL$ cannot be distinguished from scale-dependence bias alone. For our free-$p$ analysis the parameters become sufficiently degenerate that we restrict to forecasting $\fNL$ and $\tauNL$ only. In this limit constraints on $\fNL$ can be thought of as representing constraints on an effective amplitude that combines LPnG and tracer-dependent bias parameters, such as the $\mathcal{B}$ parameter in Eq.~(\ref{generalisedbeta}), while $(1+\xi^2) = \tauNL/(\frac{6}{5}\fNL)^2$ should, in principle, still be constrained even with completely free $\beta_f$. Throughout our analysis, we use $\delta_{c} = 1.42$ because it shows better agreement between the above fitting formulas and N-body simulations~\cite{KendrickMSmith_2012}.

The galaxy power spectrum observed by a given galaxy survey differs from the theoretical galaxy power spectrum modelled according to  Eq.~\eqref{Pgg_theory} in several ways. The peculiar velocities of galaxies along the line-of-sight break the isotropy of the matter power spectrum and cause the observed galaxy power spectrum to only be isotropic around the line of sight. This effect is known as redshift-space distortion and can be modelled at the linear and mildly non-linear scales relevant for the SPHEREx survey by making the replacement $b_{g,i} \rightarrow b_{g,i} + f(z)\mu^2$ and $b_{g,j}\rightarrow b_{g,j} + f(z)\mu^2$ in Eq.~\eqref{Pgg_theory}; where $f(z)$ is the linear growth rate of small-scale matter density perturbations $f(z) = d\log\delta_{cb}/d\log a$, and $\mu$ is the cosine of the angle between the wave vector $\vec{k}$ and the line of sight (this is the well-known Kaiser term~\cite{Kaiser:1987qv}). In addition to redshift-space distortion effects, there are additional multiplicative factors needed to map Eq.~\eqref{Pgg_theory} to a model of the observed galaxy power spectra. Our final expression for the observed galaxy power spectrum between the $i^{th}$ and $j^{th}$ galaxy samples of SPHEREx is (see Ref.~\cite{Shiveshwarkar:2023xjv})
\be
    \label{Pgg_obs}
    P_{ij} = f_{BF}\times (f_{\sigma_{z,i}}f_{\sigma_{z,j}})^{1/2}\times f_{AP} \large[(b_{g,i}+f\mu^2)(b_{g,j}+f\mu^2) P_{cc}+ \xi^2\Delta b_{NG,i}\Delta b_{NG,j} P_{cc} \large]\ ,
\ee
where $f_{AP}(z)$ denotes the contribution of the Alcock-Paczynski effect and $f_{BF}(k,z)$ encodes the effect of non-linear bulk flows. The Alcock-Paczynski term corrects for the error in the galaxy power spectrum measured from data assuming a fiducial cosmology which is different from the true cosmology. The term $f_{AP}(z)$ relates the comoving volume elements in the fiducial and true cosmologies and if given by :
\begin{eqnarray}
\label{f_AP}
f_{AP}(z) = \frac{H(z)D^{\text{fid}}_{A}(z)^{2}}{H^{\text{fid}}(z)D_{A}(z)^{2}}\ .
\end{eqnarray}
The term $f_{BF}(k,z)$ models the smearing of (small-scale) BAO features in the galaxy power spectrum due to the effect of non-linear bulk flows~\cite{Dore:2014cca}. Following Ref. \cite{Dore:2014cca}, 
\begin{eqnarray}
\label{bulk_flows}
f_{BF}(k,z) = \exp\left(-\frac{1}{2}k^{2}\Sigma_{\perp}^{2}-\frac{1}{2}k^{2}\mu^{2}(\Sigma_{||}^{2}-\Sigma_{\perp}^{2})\right)\ .
\end{eqnarray}
The Lagrangian displacement fields $\Sigma_{\perp}$ and $\Sigma_{||}$ are given by \cite{Dore:2014cca}
\begin{eqnarray}
\label{eq:sigma1}
\Sigma_{\perp}(z) &=& c_{\text{rec}}D(z)\Sigma_{0}\ ,\\
\label{eq:sigma2}
\Sigma_{||}(z) &=& c_{\text{rec}}D(z)(1+f(z))\Sigma_{0}\ ,
\end{eqnarray}
where $D(z)$ is the linear growth factor and $f(z)$ is the linear growth rate. We set the parameter $c_{\text{rec}}=0.5$~\cite{Dore:2014cca} and $\Sigma_{0} = 11\ \text{h/Mpc}$ for a fiducial $\sigma_{8} = 0.8$.
 The terms $f_{\sigma_{z,i}}(k,z),\ f_{\sigma_{z,j}}(k,z)$ are exponential damping factors that suppress the observed galaxy power spectrum signal at small scales along the line of sight due to the uncertainty in the determination of galaxy redshifts (namely $\sigma_{z,i}\ \text{and}\ \sigma_{z,j}$ which denote redshift uncertainties in the $i^{th}$ and $j^{th}$ galaxy populations to be observed by SPHEREx). For the $i^{th}$ SPHEREx galaxy sample, $f_{\sigma_{z,i}}(k,z) = \exp(-k^{2}\mu^2\sigma_{z,i}^{2}/H(z)^{2})$, where $H(z)$ is the Hubble rate at redshift $z$ and $\sigma_{z,i}$ is the redshift uncertainty of the $i^{th}$ galaxy sample.  We refer the reader to Refs.~\cite{Shiveshwarkar:2023xjv,Dore:2014cca} for a detailed discussion of these additional terms and their importance in the context of the SPHEREx all-sky survey. 

Proceeding further, we treat the observed linear galaxy number density fluctuations at a given (fiducial) \textbf{k}-mode and redshift as Gaussian random variables~\cite{Shiveshwarkar:2023xjv} with covariance (assuming Poisson shot noise) given by 
\begin{eqnarray}
    \label{Cov_formula}
    \langle \delta^{\dagger}_{g,i}\delta_{g,j}\rangle = P_{ij}(\vec{k},z) + \frac{\delta^{K}_{ij}}{\overline{n}_{g,i}(z)}\ ,
\end{eqnarray}
where $P_{ij}$ denotes the observed the cross-power spectrum of the $i^{th}$ and $j^{th}$ galaxy populations of SPHEREx (see Eq.~\eqref{Pgg_obs}) and $\overline{n}_{g,i}$ is the mean number density of observed galaxies in the $i^{th}$ galaxy population of SPHEREx at redshift $z$. This translates into an exponential likelihood for the galaxy power spectra at a given \textbf{k}-mode and redshift given by 
\begin{eqnarray}
\label{lkl1}
    \mathcal{L}_{\textbf{k},z} = \mathcal{J} \frac{1}{\det\mathcal{C}}\cdot\exp\left[-\text{Tr}\left(\mathcal{C}^{-1}\mathcal{D}\right)\right]\ ,
\end{eqnarray}
where the proportionality constant $\mathcal{J}$ is independent of the sampled cosmological parameters. In Eq.~\eqref{lkl1}, both $\mathcal{C}$ and $\mathcal{D}$ are the covariance matrices of galaxy number density fluctuations (at a given \textbf{k}-mode and redshift) that respectively depend on the sampled, $^s$, and fiducial, $^{fid}$, cosmological parameters in an MCMC run: 
\begin{eqnarray}
    \mathcal{C}_{ij} &=& P^{s}_{ij} + \frac{\delta^{K}_{ij}}{\overline{n}_{g,i}}\ ,\\
    \mathcal{D}_{ij} &=& P^{fid}_{ij} + \frac{\delta^{K}_{ij}}{\overline{n}_{g,i}}\ .
\end{eqnarray}

The entire SPHEREx power spectrum likelihood is the product of $\mathcal{L}_{\textbf{k},z}$ over all redshifts bins and independent \textbf{k}-modes. As stated before, the SPHEREx survey is a large-scale survey~\cite{Dore:2014cca} with the observed \textbf{k}-modes between $k_{\text{min}} = 0.001\ \text{h/Mpc}$ and $k_{\text{max}} = 0.2\ \text{h/Mpc}$ and eleven redshift bins with mean redshifts going from $\overline{z}_{\text{min}}=0.1$ to $\overline{z}_{\text{max}}=4.3$. Note that for the purposes of obtaining power-spectrum forecasts, we assume that different redshift bins and independent \textbf{k}-modes are uncorrelated -- this is a standard approximation used in forecasting the sensitivity of cosmological surveys (see Refs.~\cite{Dore:2014cca,Sprenger:2018tdb,Shiveshwarkar:2023xjv}). An MCMC run used to obtain cosmological forecasts typically maximises the likelihood by minimising $\chi^2$ defined as 
\begin{eqnarray}
    \label{chi2}
    \chi^{2} = -2\sum_{z_{i}}V(z_{i})\int_{k_{\text{min}}}^{k_{\text{max}}}\int_{-1}^{1}\frac{k^{2}dkd\mu}{2(2\pi^{2})}\log\mathcal{L}_{\vec{k},z}\ ,
\end{eqnarray}
where $V(z_{i})$ is the comoving volume surveyed in the $i^{th}$ redshift bin. 

Having thus obtained the likelihood, we use the boosted MCMC sampler MontePython v3.4 to conduct MCMC runs to fit for the following cosmological parameters: \{$\omega_{b}$ , $\omega_{cdm}$, $100\theta_s$, $A_s$, $n_s$, $z_{reio}$, $M_{\nu}$\}, as well as combinations of \{$\fNL$,  $\gNL$, $\tauNL$\}. In addition, we marginalise over the scale-independent, Gaussian parts of the galaxy biases (given by $b_{0,i} = b_{g,i}-\Delta b_{NG,i}$) as well as the bulk flow parameter $\Sigma_{0}$ which parametrises the effect of non-linear bulk flows given by $f_{BF}(k,z)$ in Eq.~\eqref{Pgg_obs} and Eqs.~(\ref{bulk_flows})--(\ref{eq:sigma2}) (see Refs.~\cite{Dore:2014cca,Shiveshwarkar:2023xjv} for more details). During our MCMC analysis, we combine our power spectrum likelihood (Eq.~\eqref{chi2},~\eqref{lkl1}) with a mock CMB power spectrum likelihood named `fake\_planck\_realistic' in MontePython -- this likelihood is designed to mimic the sensitivity of the Planck 2018 dataset in constraining cosmological parameters (see Ref.~\cite{Brinckmann:2018owf} for more details). This mock CMB likelihood does not include the CMB lensing power spectrum -- this ensures that it is not significantly correlated with our SPHEREx galaxy power spectrum likelihood. Moreover, the mock CMB likelihood does not include the effect of primordial non-Gaussianity and does not by itself provide any constraints on LPnG parameters -- this makes it equivalent to a Planck-based prior imposed in order to resolve degeneracies between $\nu\Lambda CDM$ cosmological parameters (as is done in Ref.~\cite{Dore:2014cca}).  

We use a $\nu\Lambda CDM$ cosmology with fiducial parameters $\omega_b = 0.02218$, $\omega_{\text{cdm}}=0.1205$, $100\times\theta_s = 1.04113$, $A_s = 2.032692\times 10^{-9}$, $n_{s}=0.9613$, $z_{reio} = 7.68$, $M_{\nu} \equiv \Sigma_i m_{\nu i} = 0.06$ eV. For simplicity neutrinos are implemented as three degenerate species (see, e.g, Ref.~\cite{Gerbino:2016sgw}). In this paper, we forecast constraints on non-Gaussian initial conditions with primordial trispectra that contribute comparably in magnitude to galaxy power spectra as primordial bispectra with $\fNL \sim \mathcal{O}(1)$.  To achieve this scenario in which the primordial kurtosis is comparable to the skewness  we choose fiducial values of $\fNL = 1.0$ and $\gNL = 1.0\times 10^4$. For the two-source initial conditions described in Section \ref{Theory2} that produce a $\tauNL$ not specified by $\fNL$ alone, we choose to obtain forecasts around two fiducial values of $\tauNL$, namely $\tauNL=1.3\times 10^2$ and $\tauNL = 1.3\times 10^3$, well above our fiducial value of $\fNL=1$ and consistent with CMB datasets \cite{Planck2013}.

\section{Results}
\label{sec:results}

\begin{table}
\centering
\begin{tabular}{||c|c|c||} 
 \hline
 $ p $ & $\sigma(\fNL)$ & $\sigma(\gNL)$ \\ [0.5ex] 
 \hline\hline
 1.0 & 1.06 & $0.92\times 10^{4}$ \\ [1ex]
 \hline\hline
 0.5 & 0.89 & $0.66\times 10^{4}$ \\[1ex]
 \hline
\end{tabular}
\caption{Single-parameter LPnG forecasts for $\fNL$ and $\gNL$ obtained from the SPHEREx multitracer likelihood for the $p=1$ and $p=0.5$ modelling choices (see Eq.~(\ref{betagnl}) and Eq.~(\ref{eq:betafstrong})).}
\label{control}
\end{table}

We start by presenting single-parameter forecasts for $\fNL$ and $\gNL$ which parametrise the squeezed-limit primordial bispectrum and trispectrum respectively. Table~\ref{control} shows the SPHEREx multitracer MCMC forecasts for the individual parameters $\fNL$ and $\gNL$ around their respective fiducial values $\fNL=1.0$, and $\gNL = 1.0\times 10^4$ obtained within the framework of the two-field inflationary model. Note that for the two-field inflationary model considered in this paper, it doesn't make sense to obtain individual forecasts for $\tauNL$ because $\tauNL$ cannot be non-vanishing unless $\fNL\neq 0$. The forecasts in Table~\ref{control} are obtained under strong modelling assumptions (see Section~\ref{sec:method}, Eqs.~\eqref{dbNGfNLgNL1} and~\eqref{betagnl} and surrounding text), where $\beta_{f}=\delta_{c}(b_{E}-p)$ and $\nu = (1+\delta_{c}(b_{E}-p))^{1/2}$ with $p=0.5$ or $p=1$ for all galaxy samples, yet $b_E$ is a free parameter for each galaxy population at each redshift. As expected, the forecast constraints on $\fNL$ and $\gNL$ depend on the assumed value of $p$. Our individual forecasts for $\fNL$ or $\gNL$ alone are also consistent with the observation at the end of Section~\ref{Theory1} that the contribution of $\gNL$ to the scale-dependent bias signal is suppressed w.r.t. that of $\fNL$ by four orders of magnitude, due to the relative magnitudes of the dimensionless cumulants $\langle \delta_{cb}^3\rangle/\langle\delta_{cb}^2\rangle \sim \fNL \sigma$ and $\langle \delta_{cb}^4\rangle_c/\langle\delta_{cb}^2\rangle \sim \gNL \sigma^2$ for $\fNL$ and $\gNL$ of similar size.

Having established the forecast constraints on either $\fNL$ or $\gNL$ alone, we now consider combined constraints. We conduct MCMC runs to obtain joint forecasts for the pairs of parameters $(\fNL,\gNL)$ and $(\fNL,\tauNL)$. Expanding this to include all three parameters is a challenging and computationally intensive sampling problem due to strong degeneracies between the parameters. We expect the sensitivity would be washed out and barely more illuminating than the two-parameter case, since we would only see notable constraining power for the parameter(s) with the greatest contribution to the observable signal. The exception to this is the somewhat fine-tuned case where all three parameters contribute comparably, in which case we would expect decreased sensitivity due to the strong three-way degeneracy. For the subcase with two parameters giving comparable contributions to the scale-dependent bias, we expect the result would largely asymptote to that of the two-parameter case, possibly with slightly decreased sensitivity to the parameters. Given this expectation, we do not consider it worth the significant computational resources required (and associated carbon footprint) and omit the three-parameter case.

\subsection{Joint forecasts on $\fNL$ and $\gNL$}\label{fNLgNLresults}

\begin{figure}
  \begin{center}
    \includegraphics[width=0.5\textwidth]{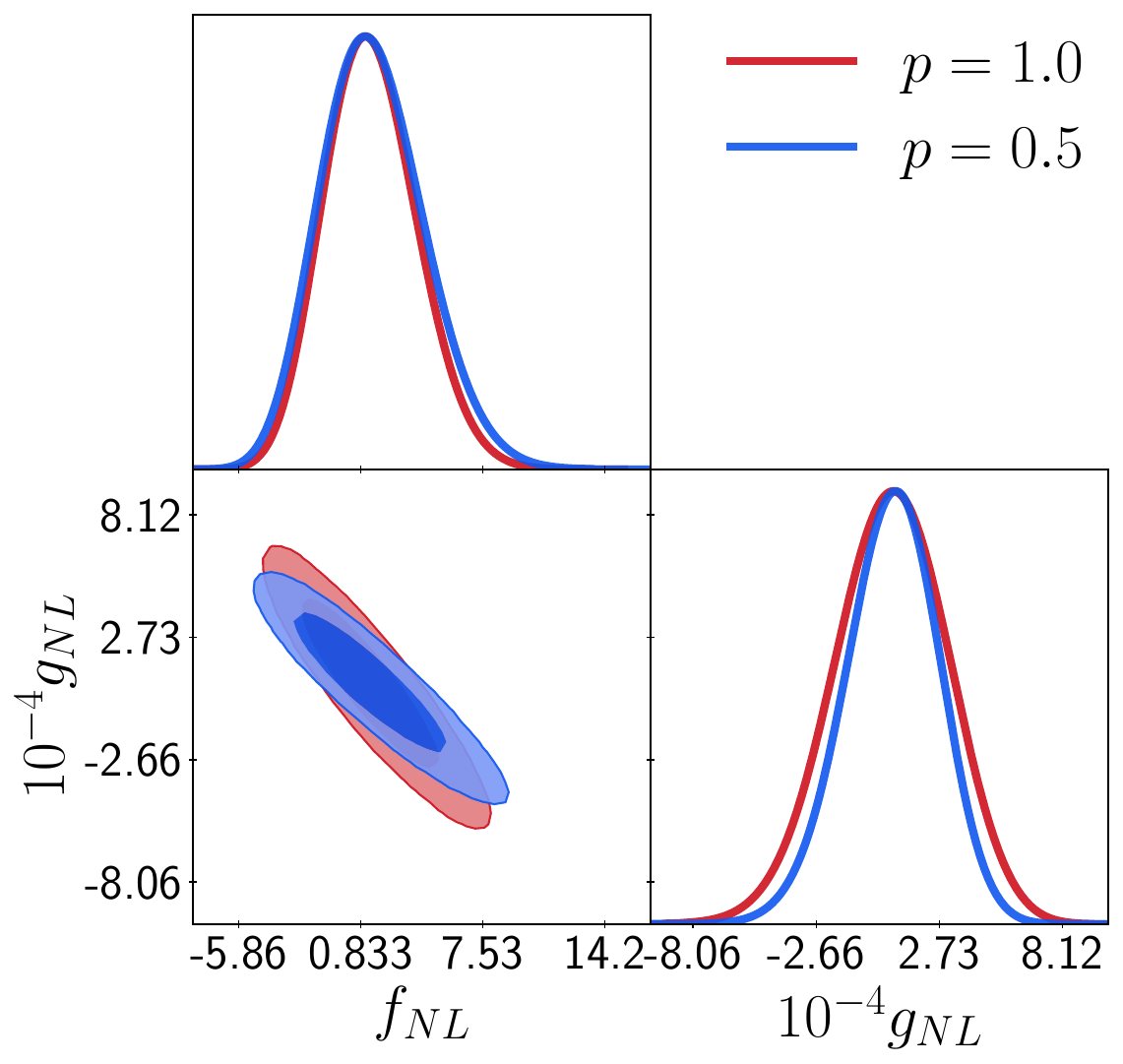}
  \end{center}
  \caption{Joint MCMC forecasts around fiducial $\fNL = 1.0$ and fiducial $\gNL = 1\times 10^{4}$ obtained from the galaxy auto- and cross-power spectra of all five SPHEREx galaxy samples. The two sets of contours illustrate constraints for two modelling choices for the scale-dependent bias, $p=1$ and $p=0.5$.}\label{fig:spherexfNLgNL}
\end{figure}

To obtain joint forecasts for $\fNL$ and $\gNL$, we conduct MCMC runs with the galaxy bias modelled as Eq.~(\eqref{dbNGfNLgNL1}), with $\beta_f$ and $\beta_g$ as in Eq.~(\ref{eq:betafstrong}) and Eq.~(\ref{betagnl}), and simultaneously fit for $\fNL$ and $\gNL$. Table~\ref{fg} shows our joint MCMC forecasts for $\fNL$ and $\gNL$ as well as their covariance around fiducial values $f_{\text{NL,fid}} = 1.0$ and $g_{\text{NL,fid}}=1\times 10^4$.  Our fiducial values for $\fNL$ and $\gNL$ are consistent with the most recent CMB constraints~\cite{Planck:2019kim} and are such that the dimensionless skewness and kurtosis of the primordial curvature ($\langle\zeta^{3}\rangle/\langle\zeta^2\rangle^{3/2}$ and $\langle\zeta^{4}\rangle/\langle\zeta^2\rangle^{2}$) are of similar order of magnitude. As mentioned in Section~\ref{Theory1}, both $\fNL$ and $\gNL$ contribute comparably to the observed signal of primordial non-Gaussianity (i.e. the scale-dependent bias) in this regime. We again consider two representative values of $p$, $p=1$ and $p=0.5$.  Controlling for the difference in the order of magnitude of $\fNL$ and $\gNL$, Table~\ref{fg} shows that we obtain nearly equal forecast constraints for $\fNL$ and $\gNL$, analogous to the individual forecasts in Table~\ref{control}. However, these forecasts are significantly weaker than their respective individual forecasts. This degradation in forecasts for $\fNL$ and $\gNL$ can be attributed to the strong degeneracy between $\fNL$ and $\gNL$, which is evident from Eq.~(\ref{dbNGfNLgNL1}). Both terms produce the same $k$-dependence in the bias, the terms differ only in how the amplitude of the $k$-dependent terms vary with tracer population through $\beta_f$ and $\beta_g$, which are themselves degenerate through the dependence on $b_E$. This degeneracy is reflected in Figure~\ref{fig:spherexfNLgNL}, which shows the joint $68\%$ and $95\%$ confidence regions for $\fNL$ and $\gNL$ obtained from our MCMC analysis. The degeneracy between $\fNL$ and $\gNL$ is also reflected in their covariance (see Table~\ref{fg}), which is $\sim -0.9$ for both the $p=1$ and $p=0.5$, with only slight variations in the covariance between the two $p$-values (visible in Figure~\ref{fig:spherexfNLgNL}) due to the small changes in the relationship between $\beta_g$ and $b_E$ in each case. This is consistent with the Fisher matrix analysis done by Ref.~\cite{Ferraro:2014jba}, where it was found that while multitracer analysis using scale-dependent halo bias can distinguish between $\fNL$ and $\gNL$, their covariance is always close to $-1$ due to their degeneracy w.r.t. the scale-dependent bias.

As shown in Table~\ref{control}, our SPHEREx multitracer forecasts under non-Gaussian initial conditions for only $\fNL \neq 0$ or only $\gNL\neq 0$ are such that $\Delta\fNL/\fNL \sim \Delta\gNL/\gNL \sim 1$ for fiducial values $\fNL=1.0$ and $\gNL=1.0\times 10^4$. This means that with systematic uncertainties eliminated and in the scenario where either $\fNL\neq 0$ or $\gNL\neq 0$, SPHEREx power spectrum measurements alone can potentially detect LPnG (at the small amplitudes here) with at most $1\sigma$ significance (with the modelling assumptions made here). It is worth noting that this continues to hold even with both $\fNL$ and $\gNL$ being as large as their fiducial values -- in fact, the prospect of SPHEREx detecting LPnG with $1\sigma$ significance is slightly improved when both are non-zero. For forecasts obtained with $p=1$, the point $\fNL=\gNL=0$ is now disfavoured with $\Delta \chi^2 = 4.6$, which puts it far outside the $68\%$ confidence region contour (as compared to the forecasts made for a single parameter in which case the point $\fNL=0$ is nearly at the boundary of the 68\% confidence interval). On the other hand, the point $\fNL=\gNL=0$ is disfavoured with $\Delta \chi^2 = 6.5 $ for forecasts obtained using the modelling assumption $p=0.5$ -- suggesting that SPHEREx power spectrum measurements alone can potentially detect LPnG at the (small) amplitudes considered here with as much as $2\sigma$ significance if SPHEREx galaxies are better characterized by $p=0.5$.\footnote{$\Delta\chi^2 = 2.3(6.17)$  for the boundary of the $68.3\% (95.4\%) $ confidence region.}

\begin{table}
\centering
\begin{tabular}{||c|c|c|c||} 
 \hline
 $ p $ & $\sigma(\fNL)$ & $\sigma(\gNL)$ & $\text{Cov}(\fNL,\gNL)$ \\ [0.5ex] 
 \hline\hline
 1.0 & 2.54 & $2.53\times 10^{4}$ & $-0.93$\\ [1ex]
 \hline\hline
 0.5 & 2.86 & $2.10\times 10^{4}$ & $-0.95$ \\[1ex]
 \hline
\end{tabular}
\caption{Joint MCMC forecast for $\fNL$ and $\gNL$ obtained from the SPHEREx multitracer likelihood for $p=1$ and $p=0.5$ modelling choices.}
\label{fg}
\end{table}

\subsection{Joint forecasts on $\fNL$ and $\tauNL$}
\label{ssec:Results_ftau}

\begin{figure}
\centering
\subfloat[Fiducial $\fNL = 1.0$ and fiducial $\tauNL = 1.3\times 10^{2}$]{\includegraphics[width=0.5\textwidth]{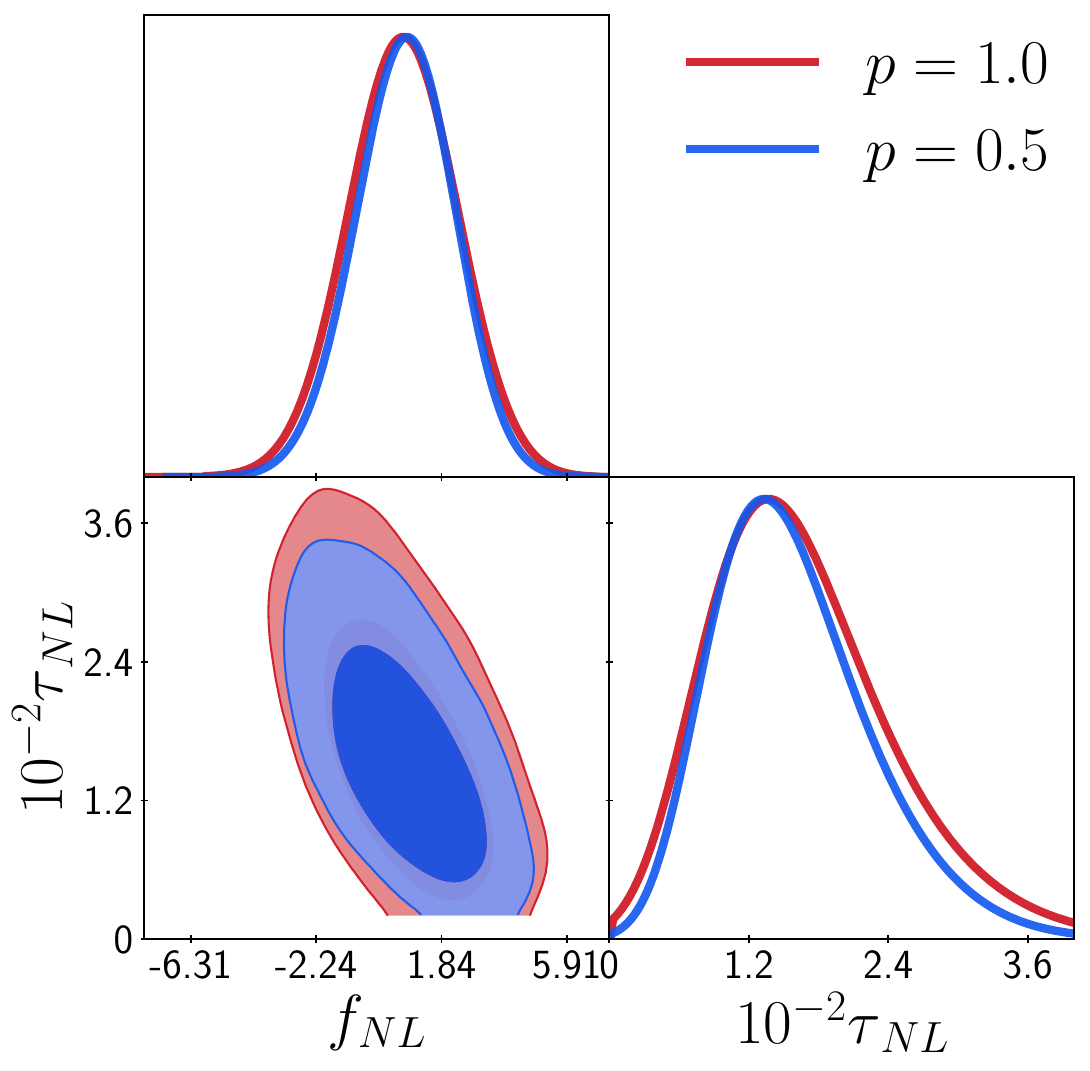}}
\subfloat[Fiducial $\fNL = 1.0$ and fiducial $\tauNL = 1.3\times 10^{3}$]{\includegraphics[width=0.5\textwidth]{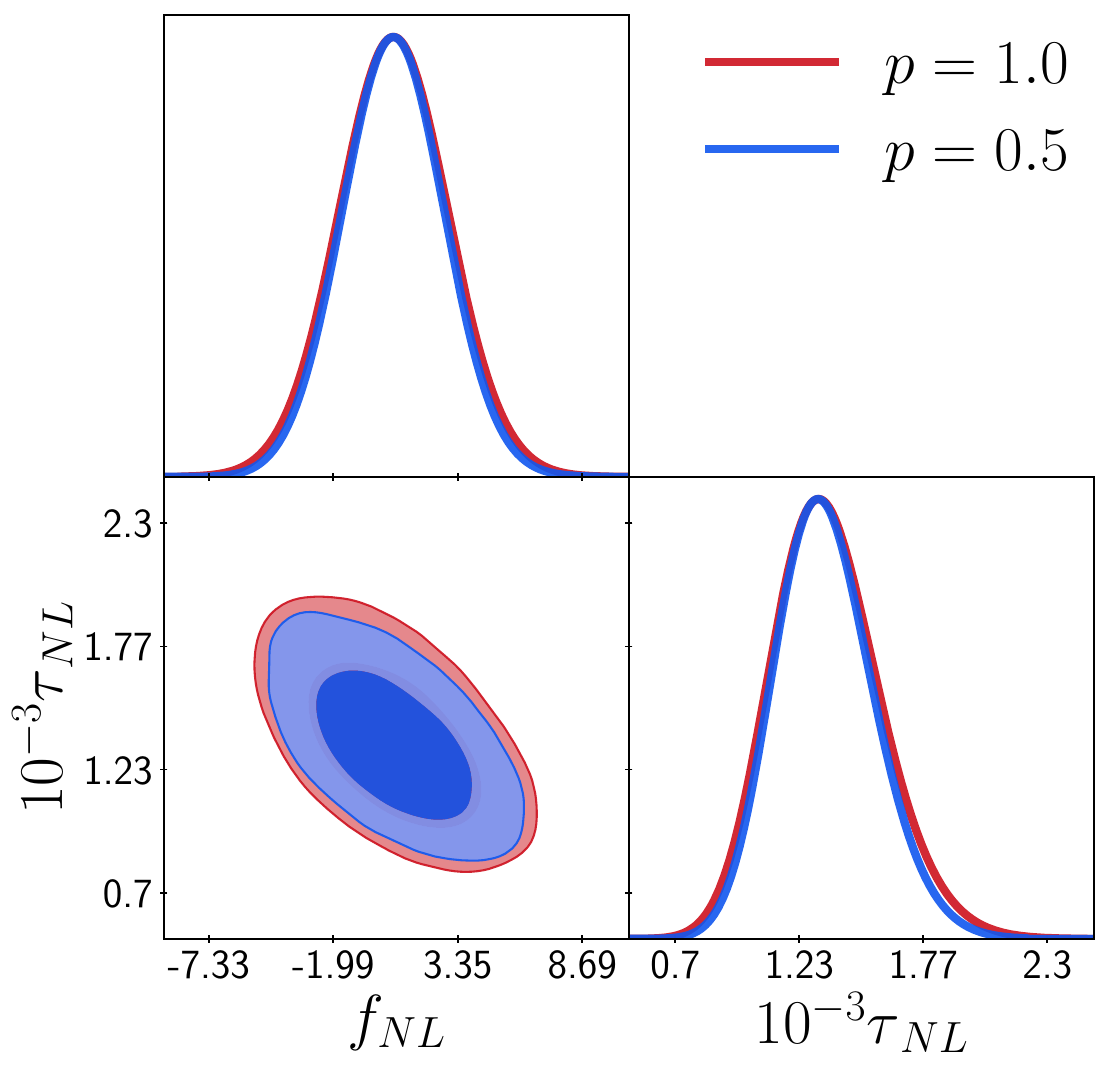}}
\caption{Joint MCMC forecasts for $\fNL$ and $\tauNL$ (obtained around the fiducial values assumed in this paper) from the scale-dependent bias of all five SPHEREx galaxy samples for $p=1$ and $p=0.5$ modelling choices.}\label{fig:spherexfNL_tauNL_forecasts}
\end{figure}

We now consider joint constraints on $\fNL$ and $\tauNL$, with $\gNL =0$. To obtain joint multitracer forecasts for $\fNL$ and $\tauNL$, we conduct two MCMC runs with fiducial values $\fNL=1,\ \tauNL=1.3\times 10^2$ and $\fNL=1,\ \tauNL=1.3\times 10^3$ respectively. As mentioned in Section~\ref{Theory}, these fiducial values are well within the current constraints, yet produce a large stochastic component of the galaxy bias. Note that while conducting our MCMC runs, we need to impose the condition $\tauNL\geq (6/5\fNL)^2$ following the Suyama-Yamaguchi inequality~\cite{SuyamaYamaguchi}.

Table~\ref{ftau} shows the joint MCMC forecast around fiducial values $\fNL=1,\ \tauNL=1\times 10^2$ and $\fNL=1,\ \tauNL=1\times 10^3$, respectively, for $p=1.0$ and $p=0.5$. Figure~\ref{fig:spherexfNL_tauNL_forecasts} shows the joint $68\%$ and $95\%$ contours for $\fNL$ and $\tauNL$ around their respective fiducial values for examples with $p=1$ and $p=0.5$. In all cases, the covariance between $\tauNL$ and $\fNL$ is $\approx -0.6$ -- this is consistent with the fact that $\fNL$ and $\tauNL$ are only partly degenerate. This is because the contribution of $\tauNL$ to the galaxy bias (unlike that of $\gNL$) goes as $\Delta b_{NG}^2$ ($\propto 1/k^4$ on large scales, see Eq.~\eqref{PggtauNL}) and does not have the same scale-dependence as the contribution of $\fNL$ ($\propto 1/k^2$ on large scales). The contour plots in Figure~\ref{fig:spherexfNL_tauNL_forecasts} as well as the covariances in Table~\ref{ftau} indeed show that while $\fNL$ and $\tauNL$ are partly degenerate, they are not as degenerate as $\fNL$ and $\gNL$. We also see that the choice of $p$ does not appear to effect the degeneracy between $\fNL$ and $\tauNL$. Comparing the left and right panels of Figure~\ref{fig:spherexfNL_tauNL_forecasts} or the two sub-tables in Table~\ref{ftau}, we see that larger fiducial values of $\tauNL$ lead to tighter constraints on $\tauNL$. This is, perhaps, unsurprising as a large $\tauNL$ should lead to a larger signal. On the other hand, we see that constraints on $\fNL$ are weakened in the presence of larger $\tauNL$. Our MCMC forecasts lead to the observation that in the absence of systematics, SPHEREx can potentially constrain large values of $\tauNL$ more precisely at the cost of degraded constraints on $\fNL$. In particular, in the scenario with $\tauNL=1\times 10^3$, SPHEREx could detect non-trivial $\tauNL$ at $\approx 5\sigma$ significance while not, at the same time, having unambiguously detected $\fNL\neq 0$. This means that SPHEREx could infer the presence of an additional light field during inflation (and thus indirectly rule out $\fNL=0$) through a robust measurement of scale-dependent \textit{stochastic} bias, if the amplitude is sufficiently large.

\begin{table}
 \subfloat[Fiducial $\fNL = 1.0$ and fiducial $\tauNL = 1.3\times 10^{2}$]{\begin{tabular}{||c|c|c|c||} 
 \hline
 $ p $ & $\sigma(\fNL)$ & $\sigma(\tauNL)$ & $\text{Cov}(\fNL,\tauNL)$ \\ [0.5ex] 
 \hline\hline
 1.0 & 1.79 & $0.78\times 10^{2}$ & $-0.59$\\ [1ex]
 \hline\hline
 0.5 & 1.64 & $0.67\times 10^{2}$ & $-0.61$ \\[1ex]
 \hline
\end{tabular}}
\quad
\subfloat[Fiducial $\fNL = 1.0$ and fiducial $\tauNL = 1.3\times 10^{3}$]{\begin{tabular}{||c|c|c|c||} 
 \hline
 $ p $ & $\sigma(\fNL)$ & $\sigma(\tauNL)$ & $\text{Cov}(\fNL,\tauNL)$ \\ [0.5ex] 
 \hline\hline
 1.0 & 2.42 & $0.24\times 10^{3}$ & $-0.56$\\ [1ex]
 \hline\hline
 0.5 & 2.22 & $0.21\times 10^{3}$ & $-0.59$ \\[1ex]
 \hline
\end{tabular}}
\caption{Joint MCMC forecast for $\fNL$ and $\tauNL$ obtained from the SPHEREx multitracer likelihood. For each fiducial value of $\tauNL$, we consider two example values of $p=1$ and $p=0.5$, entering the model for $\beta_f$ as in Eq.~(\ref{eq:betafstrong}).}\label{ftau}
\end{table}

\subsection{Impact of modelling choices}\label{ssec:modelchoices}

\begin{figure}[h!]
    \centering
    \includegraphics[width=\columnwidth]{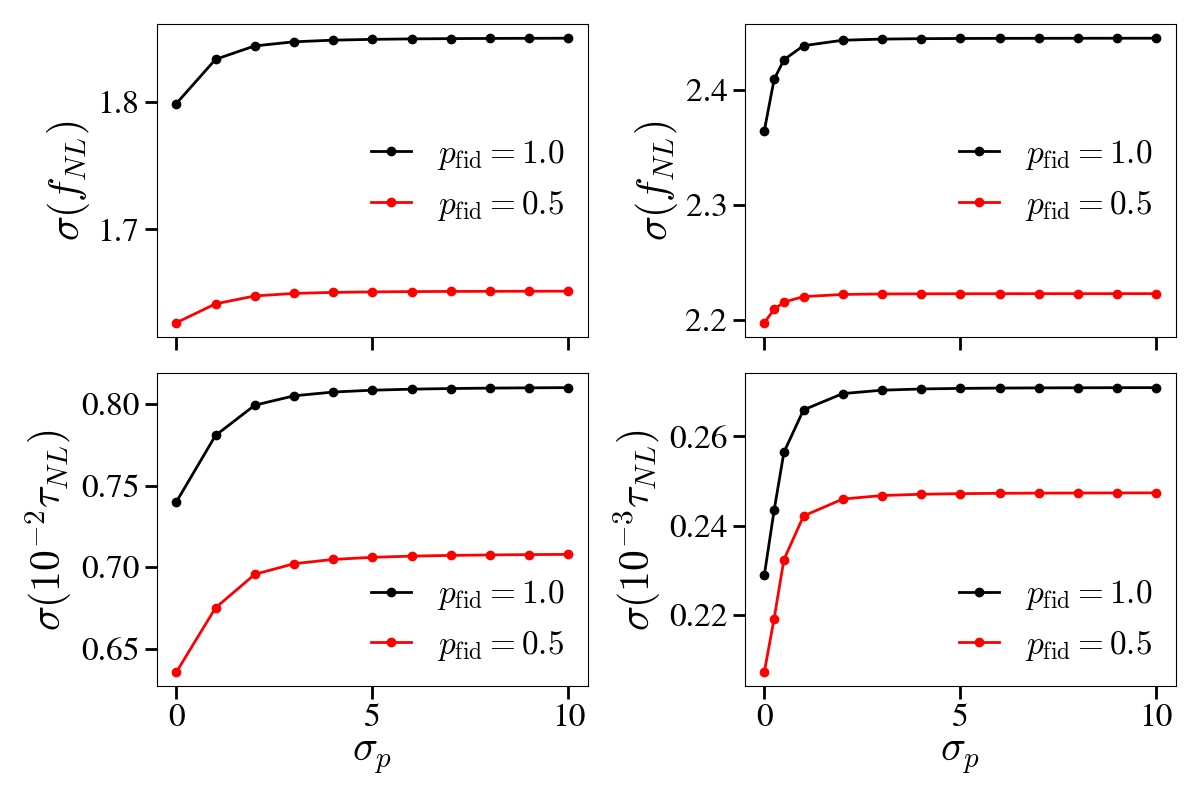}
    \caption{Joint Fisher forecasts for $\fNL$ and $\tauNL$ around $\fNL=1$ with $\tauNL=1.3\times 10^2$ (left) and $\tauNL=1.3\times 10^3$ (right) and $\beta_{f}=\delta_{c}(b_{E}-p)$ as a function of priors on $p$. The red plots show forecasts obtained with fiducial $p=0.5$ and the black plots show forecasts obtained with fiducial $p=1.0$.}
    \label{fig:sigfNLtauNL_vs_sigp}
\end{figure}

Our single-parameter, as well as joint, MCMC forecasts for $\fNL,\ \gNL,\ \text{and}\ \tauNL$ (see Tables~\ref{control},~\ref{fg},~\ref{ftau}) make it amply clear that the constraining power of any galaxy survey in constraining LPnG parameters using the scale-dependent galaxy bias varies with the modelling of the nuisance parameters $\beta_{f}$ and $\beta_{g}$ (see subsection~\ref{Theory1}). Given that $\beta_{f}$ and $\beta_{g}$ are strongly dependent on tracer population, they are difficult to model from first principles and present a big limiting factor to fully exploiting galaxy power spectrum data to constrain LPnG. The non-trivial relation between $\beta_{f}$ and the Eulerian galaxy bias with Gaussian initial conditions has been investigated with hydrodynamical separate universe simulations (see Refs.~\cite{Barreira_2020,Barreira_2022}), which provide some evidence that supports modelling $\beta_{f}$ as $\beta_{f} = \delta_{c}(b_{E}-p)$, with $p$ being a tracer-dependent number. However, modelling $\beta_{g}$ in terms of other observable parameters, such as $b_E$ is a much more intractable problem and represents a bigger modelling ambiguity. The is due both to the complicated nature of modelling $n_g$ from first principles, and the observation that even in the simplest models, $\beta_g$ depends non-linearly on $b_E$, and the $\beta_g$ for each sample will therefore depend on how tracers are selected and grouped together in data analyses \cite{KendrickMSmith_2012}.

In this paper, we have so far obtained forecasts for the constraining power of the SPHEREx galaxy surveys under the strongest modelling choices, which assume a complete knowledge of the relation between $\beta_{f},\ \beta_{g}$ and $b_E$. In particular, we assume (see Section~\ref{sec:method}) $\beta_{f}=\delta_{c}(b_{E}-p)$ and $\beta_{g}$ is given by Eq.~\eqref{betagnl} with $\nu = (1+0.5\beta_{f})^{1/2}$. Different tracer populations could have different values of $p$, which can change our MCMC forecasts proportionally (i.e. smaller p gives tighter constraints) given that the contributions of $\fNL$ and $\gNL$ are proportional to $\beta_{f}$ and $\beta_{g}$ respectively. We have so far presented constraints for $p=1.0$ and $p=0.5$ as representative values that reflect the universal halo mass function assumption and some results of hydrodynamical separate universe simulations, respectively. A more conservative approach is to make $\beta_{f}$ and $\beta_{g}$ free parameters and impose priors on them in our analysis. It is important to marginalise over the nuisance parameters $\beta_{f}$ and $\beta_{f}$ to obtain a more complete picture of how well a given galaxy survey can constrain LPnG using the galaxy power spectrum. We will now discuss how generalising modelling assumptions on $\beta_f$ and $\beta_g$ affects our forecast constraints.

If $\beta_f$ and $\beta_g$ are free parameters, Eq.~(\ref{dbNGfNLgNL1}) illustrates that they are completely indistinguishable from scale-dependent bias measurements alone. In other words, without predictions for the values of $\beta_f$ and $\beta_g$, the galaxy power spectrum loses its utility as a signal that can potentially distinguish between a primordial bispectrum and trispectrum. This situation is not all bad, however: it is still true that in the presence of squeezed-limits of higher-point primordial correlation functions, SPHEREx could potentially detect a non-trivial LPnG at high significance, yet with uncertainty in {\em which} squeezed-limit correlation functions were contributing. This is because the scale-dependent bias measurement is a measurement of a redshift-dependent weighted combination of the parameters $\fNL$, $\gNL$ together with all the higher order coefficients in the expansion (see Eq.~\eqref{zetaNG2}). Any measurement of $\fNL$ from observations of the scale-dependent bias of galaxies should therefore be interpreted as a measurement of a weighted combination of $\fNL$ and $\gNL$ (and potentially other higher-order squeezed-limit non-Gaussianity parameters) that depends on the observed tracer populations via the true values of $\beta_{f}$ and $\beta_{g}$ (as well as their counterparts for higher-order squeezed limit non-Gaussianity parameters). In particular, non-trivial squeezed limits of higher-point primordial correlation functions can potentially contribute to increasing the observable signal of LPnG -- our results show that this happens around the fiducial values of $\fNL$ and $\gNL$ consistent with current constraints from CMB datasets chosen in this paper.\footnote{Note that in principle, it is possible for $\fNL$ and $\gNL$ to be of opposite signs and be tuned such that their respective contributions to the scale-dependent bias are (nearly) equal and of opposite signs -- this could reduce the detectability of LPnG through the scale-dependent bias, though the cancellation between the parameters would vary with the values of $\beta_f$ and $\beta_g$, and thereby vary with the tracer population.} 

In the scenario in which multiple higher-order statistics contribute to scale-dependent bias, one should then consider constraining an effective amplitude of LPnG determined by $\mathcal{B}(z)$ in Eq.~(\ref{generalisedbeta}). If an effective amplitude $\mathcal{B}(z)$ is determined to be non-zero, this would already definitively prove the existence of LPnG and thereby additional light fields during inflation -- even if the individual coefficients in Eq.~(\ref{fNLgNLdef}) were undetermined. The challenge is that $\mathcal{B}(z)$ depends on both the primordial non-Gaussian parameters and the tracer populations. An individual measurement of $\mathcal{B}$ from one redshift and tracer population will have much lower signal-to-noise than a measurement of LPnG that combines data from all redshift bins and tracer populations. Indeed, Fisher matrix computations for the most constraining galaxy sample of SPHEREx performed using our likelihood pipeline indicate that a non-trivial LPnG (with $\fNL\sim 1$, $\gNL \sim 10^4$) cannot be ruled out with any significance by fitting for the individual $\mathcal{B}(z_i)$ when these parameters are free to vary \textit{independently} at each redshift $z_i$. Consequently, one needs a model for $\mathcal{B}(z)$ -- for instance, a model relating $\mathcal{B}(z)$  to the galaxy bias $b_E(z)$ or a parameterization of the redshift dependence of $\mathcal{B}(z)$ -- to be able to harness the full power of the data to unambiguously detect LPnG at these threshold amplitudes.

Despite the challenge of forecasting simultaneous constraints on $\fNL$ and $\gNL$ (and other higher-order non-Gaussian parameters) with more conservative modelling assumptions, one can move forward with the simpler $\fNL$, $\tauNL$ scenario. In this case, we consider $\gNL$ to be small in comparison to $\sim 10^4$, so that we can neglect that contribution to $\Delta b_{NG}$ and proceed with forecasting constraints on $\fNL$ and $\tauNL$ alone with relaxed assumptions about the form of $\beta_f$. To assess the impact of modelling ambiguities on our joint forecasts of $\fNL$ and $\tauNL$, we obtained Fisher forecasts\footnote{Although an MCMC analysis is more precise than a Fisher matrix analysis, we find (throughout our analysis) that the results of MCMC runs are in good agreement with the results of Fisher forecasts.} for the single LPnG parameter $\fNL$ as well as joint fisher forecasts for the pair of parameters $(\fNL,\ \tauNL)$ under a more conservative modelling assumption whereby $\beta_{f} = \delta_{c}(b_{E}-p)$ with $p$ now being a free parameter for each galaxy sample of SPHEREx. This is still a more restrictive choice than allowing $\beta_f$ (or nearly equivalently, $p$) for each galaxy sample and redshift to be a free parameter, but is supported by evidence from separate universe hydrodynamical simulations~\cite{Barreira_2020,Barreira_2022}. Moreover, this introduces only five additional parameters (one for each galaxy sample of SPHEREx).  For simplicity, we assumed the same fiducial values for each galaxy sample (namely $p=1$ and $p=0.5$) and obtained joint Fisher forecasts for $\fNL$ and $\tauNL$ (around the fiducial values assumed in this paper) using our SPHEREx likelihood with varying priors on $p$. This approach illustrates the impact of modelling choices on joint constraints on $\fNL$ and $\tauNL$. While the fiducial values of $p$ may in general different for the different galaxy samples, we note that SPHEREx power spectrum constraints on $\fNL$ are dominated by a single galaxy sample \cite{Shiveshwarkar:2023xjv}. This simple exercise should therefore help us understand how the forecasts obtained with the modelling choices in this paper could change under more conservative modelling assumptions.  

Figure~\ref{fig:sigfNLtauNL_vs_sigp} shows the joint Fisher forecasts for $\fNL$ and $\tauNL$ as function of the prior uncertainty on $p$ around the fiducial values $\fNL=1.0,\ \tauNL=1.3\times 10^2$ (left panel of Figure~\ref{fig:sigfNLtauNL_vs_sigp}) and $\fNL=1.0\ \tauNL=1.3\times 10^3$ (right panel of Figure~\ref{fig:sigfNLtauNL_vs_sigp}). Note that these Fisher forecasts are obtained with an independent Gaussian prior on each $p$ (i.e. one for each galaxy population of SPHEREx) whose width goes from $\sigma(p)=0$ to $\sigma(p)=10$. From both Figure~\ref{fig:sigfNLtauNL_vs_sigp}, we see that imposing more and more liberal priors on $p$ (and thus making the modelling of $\beta_{f}$ more ambiguous) does indeed worsen Fisher forecasts. However, their difference from Fisher forecasts obtained with a fixed value of $p$ (i.e. $\sigma(p)=0$) is $\lesssim 20\%$. The point at which $\sigma_{\fNL}$ ceases to grow, and the value at which the degradation of constraints on $\sigma_{\fNL}$ saturates, is dependent on the tracer population through fiducial values of $\beta_f$. In Figure~\ref{fig:sigfNLtauNL_vs_sigp} this is evident from the slight differences between the shapes of curves with different values of $p$. Since we observe a close agreement (within a few percent) between our MCMC analysis and the corresponding Fisher forecasts (both for fixed $p$), Figure~\ref{fig:sigfNLtauNL_vs_sigp} shows that our MCMC results (presented in Tables~\ref{ftau} and Figure~\ref{fig:spherexfNL_tauNL_forecasts}) give a good quantitative idea of how well SPHEREx can jointly constrain $\fNL$ and $\tauNL$ under more conservative modelling assumptions for the nuisance parameter $\beta_{f}$. Finally, we note again that from Eq.~(\ref{PggtauNL}) and Eq.~(\ref{eq:deltabNLC}), the ratio of $\tauNL$ to $\fNL^2$ is independent of $\beta_f$ (or $\mathcal{B}_c$) and should therefore, in principle, be measurable even with weak (or no) assumptions about those parameters. While a detection of $\xi^2$ could only be achieved if $\xi^2$ were very large (we estimate $\mathcal{O}(10^3)$) it would, independently, represent a detection of additional light degrees of freedom during inflation.

\section{Discussion} 
\label{sec:discussion}

The simplest inflationary models that generate initial conditions with LPnG have a primordial curvature perturbation which has non-Gaussianity predominantly characterised by a bispectrum (parametrised by $\fNL$) and which is sourced wholly by the fluctuations of a light field separate from the inflaton. Upcoming galaxy surveys such as SPHEREx show good promise in being able to constrain such models from observations of the galaxy power spectrum (and bispectrum) (see Ref.~\cite{Dore:2014cca}). In this paper, we analyse how well the upcoming SPHEREx all-sky survey can constrain LPnG models beyond this simplest class of models. In particular, we consider a two-field inflation model in which the primordial curvature perturbations have a non-trivial trispectrum whose squeezed limit is parameterised by $\gNL$ (in addition to a squeezed-limit bispectrum parametrised by $\fNL$). We also consider the scenario in which primordial curvature fluctuations are sourced from fluctuations of \textit{both} the inflaton and the curvaton. In this case, the contribution of inflaton fluctuations to the primordial curvature perturbations is Gaussian whereas the contribution of curvaton fluctuations shows LPnG. A substantial inflaton contribution to the primordial curvature perturbations decorrelates the collapsed limit of the primordial trispectrum from the square of the primordial bispectrum -- thus introducing an additional parameter in the model (called $\tauNL$) which parametrises the collapsed limit of the primordial trispectrum. Thus, the non-Gaussianity in this model is characterised by three parameters namely $\fNL$, $\gNL$ and $\tauNL$. In this paper, we present joint MCMC forecasts for the LPnG parameters $\fNL,\ \gNL,\ \tauNL$ from galaxy power spectra obtained in the context of the SPHEREx all-sky survey. 

In Section~\ref{Theory1}, we show that the squeezed limits of arbitrary higher-point correlation functions of the primordial curvature perturbations all impart the same characteristic scale-dependence to the bias of matter density tracers -- which is an important signal targeted by large-scale galaxy surveys that intend to constrain LPnG using galaxy power spectra. SPHEREx galaxy power spectra could therefore constrain or detect LPnG due to arbitrarily high-order correlation functions. In the context of the two-field inflation model assumed in this paper, this means that we can obtain joint MCMC forecasts for the pair of parameters $\fNL$ and $\gNL$ using the scale-dependent galaxy bias. However, it also means that $\fNL$ and $\gNL$ are strongly degenerate w.r.t. their signature on the galaxy power spectrum, causing their joint MCMC forecasts to be substantially degraded compared to MCMC forecasts for the individual parameters $\fNL$ and $\gNL$. Figure~\ref{fig:spherexfNLgNL} and Table~\ref{fg} show our joint MCMC constraints on $\fNL$ and $\gNL$ around fiducial values compatible with state-of-the-art constraints from CMB datasets. As expected, the joint forecasts on $\fNL$ and $\gNL$ are degraded compared to their single-parameter forecasts (in Table~\ref{control}). The strong degeneracy between $\fNL$ and $\gNL$ is due to the fact that their contributions to $\Delta b_{NG}$ have the same $k$-dependence and differ only in their dependence on the Eulerian bias, $b_E$. This degeneracy is reflected in their covariance, $\sim -0.9$, which also depends on modelling choices for $\Delta b_{NG}$, the scale-dependent bias (Eqs.~(\ref{dbNGfNLgNL1}),~(\ref{betagnl}), and~(\ref{eq:betafstrong}). 

Unlike $\fNL$ and $\gNL$, the parameter $\tauNL$ parametrises the collapsed limit of a primordial correlation function and leads to a stochasticity in the galaxy power spectrum which increases at the largest scales (see Section~\ref{Theory2}). Figure~\ref{fig:spherexfNL_tauNL_forecasts} and Tables~\ref{ftau} show our joint MCMC forecasts for $\fNL$ and $\tauNL$ around fiducial values consistent with constraints obtained from CMB datasets. In all cases, the covariance between $\fNL$ and $\tauNL$ is $\sim -0.6$. This is a reflection of the fact that while $\fNL$ and $\tauNL$ are quite degenerate in their impact on the galaxy power spectrum, they are not as degenerate as $\fNL$ and $\gNL$ owing to the fact the $\fNL$ terms in $\Delta b_{NG}$ scale as $1/k^2$ while the $\tauNL$ terms scale as $1/k^4$. Our results also indicate that the SPHEREx survey can potentially constrain a large, but CMB-consistent, value of $\tauNL$ tightly though with degraded constraints on $\fNL$. As noted in Section~\ref{ssec:Results_ftau}, a robust measurement of $\tauNL$ would also indicate the presence of an additional light field during inflation even if the parameter $\fNL$ could not be distinguished from zero. 

In Section~\ref{ssec:modelchoices}, we note that the constraints on $\fNL$, $\gNL$ and $\tauNL$ obtainable from galaxy power spectra depend on how one chooses to model the nuisance parameters $\beta_{f}$ and $\beta_{g}$ (see Eq.~\eqref{dbNGfNLgNL1}). Modelling ambiguity in $\beta_{f}$ as well as $\beta_{g}$ reduces the utility of the galaxy power spectrum in jointly constraining $\fNL$ and $\gNL$. This is because the scale-dependent bias effectively probes a weighted (redshift-dependent) combination of the parameters $\fNL$ and $\gNL$, where the relative weights are set by $\beta_f$ and $\beta_g$. In a general LPnG scenario (e.g. Eq.~(\ref{fNLgNLdef}) to all orders in $\mathcal{Z}$), scale-dependent bias measurements can conservatively only help us ascertain the presence/absence of an additional light degree of freedom (over and above the inflaton); modelling ambiguities in the bias expansion (Eq.~(\ref{generalisedbeta}) precludes determining which non-trivial squeezed limits of higher-point functions are sourcing scale-dependent bias). To obtain model-independent constraints on different inflationary models that generate LPnG we need to exploit higher-order galaxy correlation  measurements, which provide information complementary to measurements of galaxy power spectra. 

Including the galaxy bispectrum (as well as higher-point galaxy correlation functions) in addition to the galaxy power spectrum may break the perfect degeneracy between the scale-dependent galaxy bias and the parameter $\beta_{f}$ and can tighten the constraint on $\fNL$. However, theoretically modelling galaxy bispectra introduces additional bias parameters (such as the $b_{\phi\delta}(z)$ parameters in~\cite{Cabass:2022ymb}) over and above the parameter $\beta_{f}$ defined in Eq.~\eqref{dbNGfNLgNL1} (see for example~\cite{Cabass:2022ymb}) -- this presents additional modelling challenges and may necessitate modelling assumptions which significantly affect constraints on $\fNL$, $\gNL$ and $\tauNL$\footnote{Note that $\beta_{f}$ is also present as a nuisance parameter in the modelling of galaxy bispectra~\cite{Cabass:2022ymb}. Therefore, the bispectrum also loses its utility in constraining $\fNL$ in the absence of any knowledge of $\beta_{f}$. The parameter $\beta_{f}$ is essentially the same as the parameter $b_{\phi}$ (upto a multiplicative factor) in the EFTofLSS formalism.}. Moreover, constraints on the LPnG parameter $\fNL$ obtained from the bispectrum really come from information at non-linear scales unlike constraints obtained from the scale-dependent bias which are dominated by information from linear/quasi-linear scales (see~\cite{Shiveshwarkar:2023xjv}). On the other hand, as discussed previously, our joint forecasts show that galaxy power spectra alone can provide a powerful test of single-field inflation free from such complications and more agnostic to which statistics are sourcing the scale-dependent bias. We leave a detailed analysis of the constraining power of the galaxy bispectrum (as well as the galaxy trispectrum and other higher-order galaxy correlation functions) obtained by SPHEREx in constraining LPnG to future work.

As for $\fNL$ and $\tauNL$, we note in Section~\ref{ssec:modelchoices} that our joint forecasts are altered by $\lesssim 20\%$ under more conservative modelling choices than the ones we have assumed in our primary analysis. This means that the conclusion that SPHEREx can potentially measure $\tauNL$ more precisely using galaxy power spectra at the cost of degraded constraints on $\fNL$ likely holds true even under more conservative modelling choices. Adding galaxy bispectrum data can only make the constraints better~\cite{DAmico:2022gki}.

Since modelling ambiguities represent a big limitation in using galaxy power spectrum data to constrain LPnG, it would be interesting to investigate optimal ways of mitigating their impact. In particular, the fact that they are late-time parameters and encode the non-linear physics of galaxy formation leads to the observation that they should couple to primordial non-Gaussianities as well as non-Gaussianities generated through gravitational non-linearities in the same way. This approach might help suggest novel ways to break their degeneracy with LPnG parameters. We plan to investigate along these lines in the future.

\acknowledgments
We acknowledge helpful conversations with Benjamin Wallisch and Massimiliano Lattanzi. Results in this paper were obtained using the high performance computing system at the Institute for Advanced Computational Science at Stony Brook University. CS is grateful for support from the NASA ATP Award 80NSSC20K0541. ML is supported by the Department of Physics and the College of Arts and Sciences at the University of Washington and by the Department of Energy grant DE-SC0023183. TB was supported by ICSC -- Centro Nazionale di Ricerca in High Performance 18 Computing, Big Data and Quantum Computing, funded by European Union -- NextGenerationEU.

\newpage
\bibliographystyle{JHEP}
\bibliography{references.bib}

\end{document}